\documentclass[pdftex,number,12pt,a4paper]{article}

\usepackage{multirow}
\usepackage{authblk}

\usepackage{graphicx}
\usepackage[squaren]{SIunits}
\usepackage{color}

\def\LAT{the {\it Fermi}-LAT}

\def\plotdir{.}

\begin{document}

\title{Performance measurement of HARPO: a Time Projection Chamber as a gamma-ray telescope and polarimeter}

\author[1]{P.~Gros\thanks{philippe.gros at in2p3.fr}}
\author[2]{S.~Amano}
\author[3]{D.~Atti\'e}
\author[3]{P.~Baron}
\author[3]{D.~Baudin}
\author[1]{D.~Bernard}
\author[1]{P.~Bruel}
\author[3]{D.~Calvet}
\author[3]{ P.~Colas}
\author[4]{S.~Dat\'e}
\author[3]{A.~Delbart}
\author[1]{M.~Frotin\footnote{Now at  GEPI, Observatoire de Paris, CNRS, Univ. Paris Diderot, Place Jules Janssen, 92190 Meudon, France}}
\author[1]{Y.~Geerebaert}
\author[1]{B.~Giebels}
\author[5]{D.~G\"otz}
\author[2]{S.~Hashimoto}
\author[1]{ D.~Horan}
\author[2]{T.~Kotaka}
\author[1]{M.~Louzir}
\author[1]{F.~Magniette}
\author[2]{Y.~Minamiyama}
\author[2]{ S.~Miyamoto}
\author[4]{H.~Ohkuma}
\author[1]{P.~Poilleux}
\author[1]{I.~Semeniouk}
\author[3]{P.~Sizun}
\author[2]{ A.~Takemoto}
\author[2]{M.~Yamaguchi}
\author[3]{R.~Yonamine}
\author[1]{S.~Wang\footnote{Now at INPAC and Department of Physics and Astronomy, Shanghai Jiao Tong University, Shanghai Laboratory for Particle Physics and Cosmology, Shanghai 200240, China}}

\affil[1]{LLR, Ecole Polytechnique, CNRS/IN2P3, 91128 Palaiseau, France}
\affil[2]{LASTI, University of Hy\^ogo, Japan}
\affil[3]{CEA/Irfu Universit\'e Paris Saclay, France}
\affil[4]{JASRI/SPring8, Japan}
\affil[5]{AIM, CEA/DSM-CNRS-Universit\'e Paris Diderot, France}

\maketitle

\begin{abstract}
We analyse the performance of a gas time projection chamber (TPC) as a high-performance gamma-ray telescope and polarimeter in the e$^+$e$^-$ pair-creation regime.
We use data collected at a gamma-ray beam of known polarisation.
The TPC provides two orthogonal projections $(x,z)$ and $(y,z)$ of the tracks induced by each conversion in the gas volume.
We use a simple vertex finder in which vertices and pseudo-tracks exiting from them are identified.

We study the various contributions to the single-photon angular resolution using Monte Carlo simulations, compare them with the experimental data and find that they are in excellent agreement.
The distribution of the azimuthal angle of pair conversions shows a bias due to the non-cylindrical-symmetric structure of the detector.
This bias would average out for a long duration exposure on a space mission, but for this pencil-beam characterisation we have ensured its accurate simulation by a double systematics-control  scheme, data taking with the detector rotated at several angles with respect to the beam polarisation direction and systematics control with a non-polarised beam.

We measure, for the first time, the polarisation asymmetry of a linearly polarised gamma-ray beam in the low energy pair-creation regime.
This sub-GeV energy range is critical for cosmic sources as their spectra are power laws which fall quickly as a function of increasing energy.

This work could pave the way to extending polarised gamma-ray astronomy beyond the MeV energy regime.
\end{abstract}

\section{Introduction}

A number of groups are developing pair-conversion detector technologies alternative to the tungsten-converter / thin-sensitive-layer stacks of the COS-B / EGRET / AGILE / {\it Fermi}-LAT series, to improve the single-photon angular resolution.
Presently, observers are almost blind in the 1-100\,\mega\electronvolt\ energy range, mainly due to the degradation of the angular resolution of e$^+$e$^-$ pair telescopes at low energies: to a large extent, the sensitivity-gap problem is an angular-resolution issue~\cite{McEnery:eASTROGAM}.

We have shown~\cite{Bernard:2012uf} that gaseous detectors, such as TPCs (time projection chambers), can enable an improvement of up to one order of magnitude in the single-photon angular resolution (0.5\degree at 100\,\mega\electronvolt) with respect to \LAT\ (5\degree at 100\,\mega\electronvolt), a factor of three better than what can be expected for silicon detectors (1.0-1.5\degree @ 100\,\mega\electronvolt). 
With such a good angular resolution, and despite a lower sensitive mass, a TPC can close the sensitivity gap at the level of $10^{-6}$\,\mega\electronvolt\per\centi\meter\squared\second) between 3 and 300\,\mega\electronvolt. 
In addition, the single-track angular resolution is so good that the linear polarisation fraction 
can be measured.

We first give a brief overview of the experimental configuration, as well as the simulation of the detector and the event reconstruction.
We then describe the analysis procedure, and in particular the event selection.
Finally, we show the measured performance of the detector, in terms of angular resolution and polarimetry.
The difficulties encountered and the potential for improvement are discussed.

\section{$\gamma$-ray astronomy and polarimetry with a TPC}

TPCs are simple and robust particle detectors widely used
in high-energy physics \cite{Attie:2009zz}.
A volume of matter is immersed into an electric field, so that the
ionisation electrons produced by the passage of high-energy charged particles
drift and are collected on an anode plane.
The anode is segmented so as to provide a 2D image of the electrons raining on it as a function of drift time.
The measurement of the drift time provides the third coordinate.
In our case, the field is uniform, so that the electron
trajectories are straight lines and the electron drift velocity is
constant and uniform.
Noble gases (mainly helium to xenon) are very convenient as they allow
free electrons to drift freely over long distances.

In our case the TPC is used as an active target, that is at the same
time the converter in which the $\gamma$-ray converts and the tracker
in which the two lepton trajectories are measured, a situation that induces
conflicting constraints: for a given volume one would want to increase
the matter density and the noble gas atomic number $Z$, so as to increase the
telescope effective area (Fig. 7 of Ref. \cite{Bernard:2012uf}) 
but in so doing, the single-track angular resolution and therefore the
single-photon angular resolution would degrade (Fig. 6 of
Ref. \cite{Bernard:2012uf}).
For a liquid xenon TPC, for example, the angular resolution would show
no improvement with respect to that of \LAT.
When the effective area and the angular resolution are combined, the
point-like source sensitivity turns out to be barely affected by the
gas choice (Fig. 8 right of Ref. \cite{Bernard:2012uf}, estimated for
a given gas mass of $10\,\kilo\gram$).

When one considers in addition polarimetry, the measurement of the
linear polarisation fraction $P$ and angle $\phi_0$ of the incoming
radiation, the issue is the measurement of the azimuthal angle of the
final state leptons before multiple scattering in the tracker blurs
that information: the use of a liquid or solid TPC would need the
tracking of sub-millimeter-long track segments, which is out of reach: we
are bound to use a gas TPC (section 5 of Ref. \cite{Bernard:2013jea}).
Here again we see the same conflicting effects at work.
Increasing $Z$ and / or the pressure increases the $\gamma$-ray statistics
but also degrades the dilution factor due to multiple scattering
(Fig. 26 of Ref. \cite{Bernard:2013jea}).
For a $1\,\meter^3$ $5\,\bbar$ argon detector exposed for one year,
full time with a perfect efficiency, the precision of the measurement
of $P$ for a bright source such as the Crab pulsar is expected to be
of $\approx 1.4 \%$ (including experimental cuts), that is a
$5\,\sigma$ MDP (minimum detectable polarisation) of $\approx 7 \%$.

A multi-atomic component (means here $n>2$) added to the gas
absorbs the U.V. photons created in the amplification process before
they can reach the cathode and induce deleterious or even lethal
sustained discharges, hence the name ``quencher''.
Also electrons accelerated during their drift lose energy by inelastic
collisions with the quencher molecules, they cool down, which
mitigates the diffusion strongly.
A large choice of possible quenchers is available, including the
alcanes \cite{Attie:2009zz}.

In our case of a $2.1\,\bbar$ argon-based gas mixture, the
active-target radiation length $X_0$ is of about $56\,\meter$, so 
the probability of the pair conversion of a given photon crossing the
detector is low.
The effective area is  proportional to the gas mass, while for a
thick detector it is proportional to the geometrical surface.
The probability of photon ``loss''  by Compton scattering is small
too and the various possible processes are not competing with each
other: the pair/Compton cross section ratio is irrelevant here, in
contrast to thick detectors.

Astronomers also need to measure the energy of the collected photons,
that is, here, the momentum of each track.
This can be performed by a number of techniques including calorimetry,
magnetic spectrometry, transition radiation detection, all of which
would present a challenge to the mass budget on a space mission for
the (multi)-cubic-meter detector that would make the desired
effective area possible.
An other method uses the track-momentum dependence of multiple
scattering to obtain a measurement of the momentum from an analysis of
the angle deflections in the TPC itself \cite{Moliere}.
A pending question was the optimisation of the longitudinal
segmentation pitch over which these deflections are computed
(section 6 of Ref. \cite{Bernard:2012uf}).
An optimal treatment has been obtained recently, by a Bayesian
analysis of the filtering innovations of a series of Kalman filters
applied to the track \cite{Frosini:2017ft}:
meaningful results can be obtained with a gas TPC below
$100\,\mega\electronvolt/c$ (Fig. 11 of Ref.  \cite{Frosini:2017ft}).
  
\section{Experimental setup}

The HARPO (Hermetic ARgon POlarimeter) detector~\cite{Gros:TIPP:2014} is a demonstrator of the performance of a TPC for measuring polarised $\gamma$ rays.
It was designed for a validation on the ground in a photon beam.
The most critical constraints related to space operation were taken into account, such as the reduced number of electronic channels and long-term gas-quality preservation~\cite{Frotin:2015mir}.
It comprises a (30\,\centi\metre)$^3$ cubic TPC, designed to use a noble gas mixture from 1 to 4\,\bbar.
The present work uses an Ar:isobutane (95:5) gas mixture at 2.1\,\bbar.
A drift cage provides a 220\,\volt\per\centi\metre\ drift field.
The electrons produced by the ionisation of the gas drift along the electric field toward the readout plane at a constant velocity $v_{\rm drift} \approx 3.3\,\centi\metre\per\micro\second$.
The readout plane is equipped with two Gas Electron Multipliers (GEMs)~\cite{GEM} and one Micromesh Gas Structure (Micromegas)~\cite{micromegas} to multiply the electrons.
The amplified electrons' signal is collected by two sets of perpendicular strips at a pitch of $1\,\milli\meter$ (regular strips in the $X$-direction, and pads connected together by an underlying strip in the $Y$-direction, see Fig.~4 of Ref. \cite{Bernard:2012jy}).
The signals are read out and digitised with a set of AFTER chips~\cite{Calvet2014zva} and the associated Front End Cards (FECs).

Even though a cubic structure cannot be expected to behave as a fully isotropic detector, efforts have been made to have the longitudinal properties of the TPC (along $Z$) similar to the transverse ones ($X, Y$).
The time sampling was set to $30\,\nano\second$ so that, given the electron drift velocity, the TPC longitudinal sampling was close to the transverse (strip) pitch.
The main residual difference are the (longitudinal) shaping of the electronics, of $100\,\nano\second$ RMS, and an un-anticipated saturation of the electronics preamplifier that affected one channel (strip) independently from the others.

The transverse diffusion coefficient for that gas at that pressure was
of about $380\,\micro\meter/\sqrt{\centi\metre}$, which makes the
pitch size not far from optimal over most of the drift length range
(see Fig. 7 of Ref.  \cite{Arogancia:2007pt});  the longitudinal
coefficient was of about $220\,\micro\meter/\sqrt{\centi\metre}$.

The HARPO TPC was set up in the NewSUBARU polarised photon beam line~\cite{Horikawa2010209} in November 2014.
The photon beam is produced by Laser Compton Scattering (LCS) of an optical laser on a high energy (0.6-1.5\,\giga\electronvolt) electron beam.
A lay-out of the experiment can be found in Fig. 2 right of Ref. \cite{Delbart:2015rmp}.
Using lasers of various wavelengths and different electron beam energies~\cite{Wang:TPC:2015}, 13 photon energies from 1.74\,\mega\electronvolt\ to 74.3\,\mega\electronvolt\ were obtained.
A graphical representation of the $\gamma$-ray energies for which we
did take data, as a function of laser wavelength and electron beam
energy, can be found in Fig. 5 left of Ref. \cite{Delbart:2015rmp}.

The Compton edge of the laser inverse Compton scattering, that is,  the
highest part of the $\gamma$-ray energy spectrum, was selected by
collimation on axis, a lead brick with a $4\,\milli\meter$-diameter
hole located $24\,\meter$ downstream of the laser-electron interaction
point, defining a $83\,\micro\radian$ half-aperture divergence beam.
After collimation, the polarization of the laser beam is almost
entirely transferred to the $\gamma$-ray beam
(Eq. (39) and Fig. 7 of Ref. \cite{Sun:2011es}).
For the collisions of a $0.9824\,\giga\electronvolt$ electron beam
with a $1.54\,\micro\meter$ Erbium laser beam, for example, the
$\gamma$-ray beam energy and the polarisation transfer varied from
$11.8\,\mega\electronvolt$ and unity on axis to
$11.4\,\mega\electronvolt$ and 0.999 close to the collimator jaw,
respectively.

In order to mitigate systematic effects due to the geometry of the detector, the detector was rotated around the beam axis to 4 different angular positions (-45, 0, 45 and 90\,\degree).
Finally, for some configurations (in particular at low energy), data were also taken with randomly polarised photons as a reference.

A trigger system specific to the beam configuration was built using the signals from scintillators, from the micromegas and from the laser \cite{Geerebaert:2016dyv}.
A study of basic event characteristics for various trigger configuration showed that photon signals were recorded with about 50\% efficiency, while about 99\% of the background was rejected~\cite{Wang:2015thesis}.
For that data set \cite{Wang:2015thesis}, we were logging data at
about $65\,\hertz$ ($52\,\hertz$ of signal, $13\,\hertz$ of background),
that is with a deadtime of 10\%, given the digitisation time of
$1.67\,\milli\second$, while the incident background rate was of
$\approx 1.8\,\kilo\hertz$.

A specific event reconstruction procedure for pair conversion events was developed~\cite{Gros:SPIE:2016}.
The reconstruction is focused on the local properties of vertices.
It does not include any tracking, and does not give any global event information.
We show in the following how this limitation is bypassed using the characteristics of the beam configuration.


\section{Data selection}

Data were taken with 13 values of photon energy from 1.74\,\mega\electronvolt\ to 74.3\,\mega\electronvolt, but not all of them are usable for the analysis.
There are two main difficulties:
\begin{itemize}
\item Low energy points (below 4\,\mega\electronvolt) were obtained using a CO$_2$ laser without pulsing, and with a high intensity beam.
This created a lot of ``pile-up'' events, where several interactions occurred in the detector within the 15\,\micro\second\ readout window.
Besides, these low energy data are largely dominated by Compton scattering.
Data below 4\,\mega\electronvolt\ were therefore discarded.
\item The pre-amplifier in the AFTER electronics was found to saturate when a large charge accumulates on a single channel over a few dozen \micro\second.
This induced a loss of signal when tracks were aligned with the drift direction $z$ of the TPC.
This signal loss cannot be corrected, and creates a systematic bias in the data.
A simple model was used in the simulation which is sufficient to accurately reproduce the systematic bias in the data up to 20\,\mega\electronvolt.
We do not show polarimetry measurements above this energy.
\end{itemize}
\section{Event selection\label{sect:gammaSelection}}

The reconstruction~\cite{Gros:SPIE:2016} relies on local event geometry in 2 dimensions, without tracking.
The reconstructed 2D vertices can have one or two associated pseudo-tracks, to accommodate the possibility of overlapping tracks.
If both projections of the same vertex have two associated pseudo-tracks, there is a two-fold ambiguity. 
This ambiguity is resolved by using the ionisation fluctuation along the particle trajectories.
The two 2D projections are then combined to obtain a 3D picture.

A reconstructed vertex carries two pieces of information:
\begin{itemize}
\item the vertex position $\vec{x}_{v}$;
\item the direction $\vec{u}_{\pm}$ of the associated particles, which are described arbitrarily as electron (``-'') or positron (``+''). 
The sign of the charge is not necessary for either gamma-ray astronomy or gamma-ray polarimetry.
\end{itemize}
The opening angle $\theta_{+-}$ is defined as:
\begin{equation}
\theta_{+-}=\arccos{\vec{u}_{+} \cdot \vec{u}_{-} }
  .
\end{equation}
A vertex can have either one or two associated particles. 
A vertex with only one associated particle is given an opening angle $\theta_{+-}=0$.

\begin{figure}[ht]
\begin{center}
  \includegraphics[width=0.82\linewidth]{\plotdir/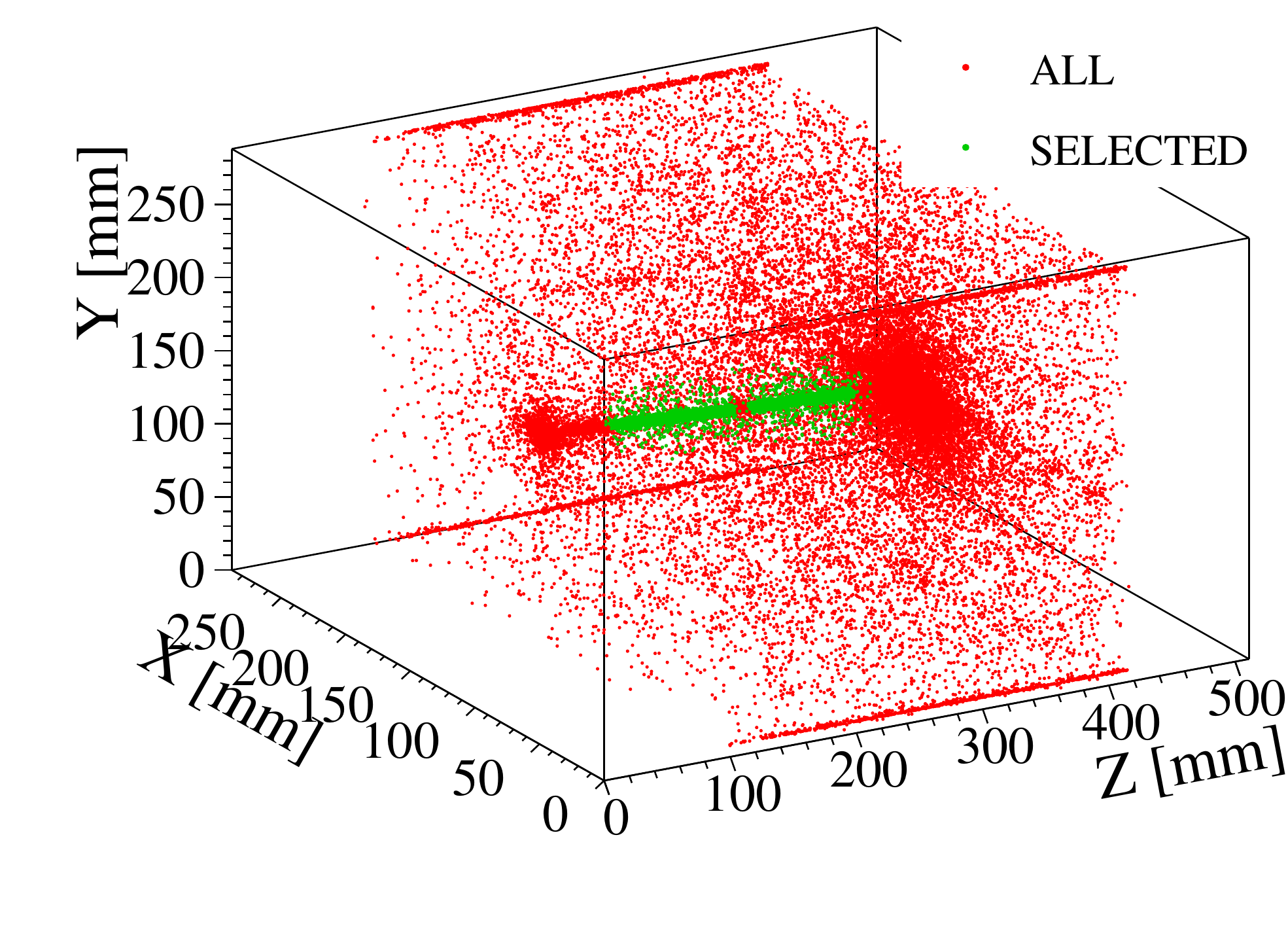}
  \caption{
    Space distribution of reconstructed vertices in the TPC.
    The high density region around $z=100\,\milli\meter$ corresponds to the entry points for conversion events in the material upstream of the gas.
    The high density region around $z=400\,\milli\meter$ corresponds to the exit points downstream.
    The data correspond to 30 minutes of data taking in the 11.8\,\mega\electronvolt\ photon beam.
    In the middle, the vertices corresponding to interactions of the photon beam with the gas are selected.
    Some effects from electronics noise appears on the edges.
    \label{fig:analyses:selection:spaceCut}
  }
\end{center}
\end{figure}

There are several reconstructed vertices for each event (in particular entry and exit points are considered as ``single particle vertices'').
Further topological information is needed to separate real vertices from background.
Using the specific configuration of the photon beam, vertices are rejected that are away from the beam or close to the walls of the TPC.
Figure~\ref{fig:analyses:selection:spaceCut} shows the space distribution of the reconstructed vertices in the TPC.
The beam region is visible as a central zone with a higher density.
The vertices around are rejected.

\begin{figure}[ht]
  \begin{center}
    \setlength{\unitlength}{0.94\textwidth}
    \begin{picture}(1,1)(0,0)
      \put(0,0.5){
        \includegraphics[width=0.5\unitlength]{\plotdir/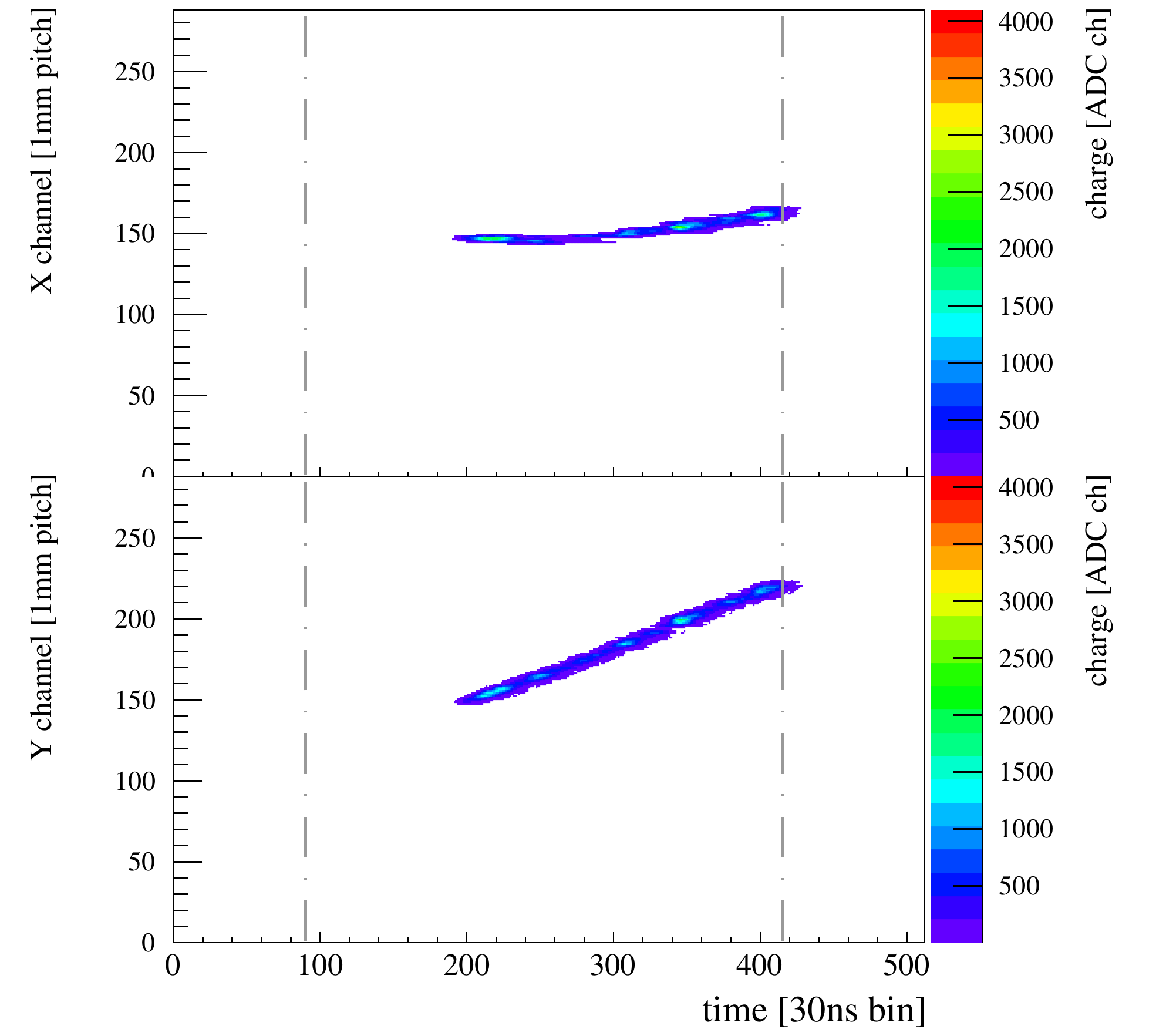}
      }
      \put(0.15,0.91){\normalsize Compton}
      \put(0.15,0.87){\normalsize 4.68\,\mega\electronvolt}
      \put(0.5,0.5){
        \includegraphics[width=0.5\unitlength]{\plotdir/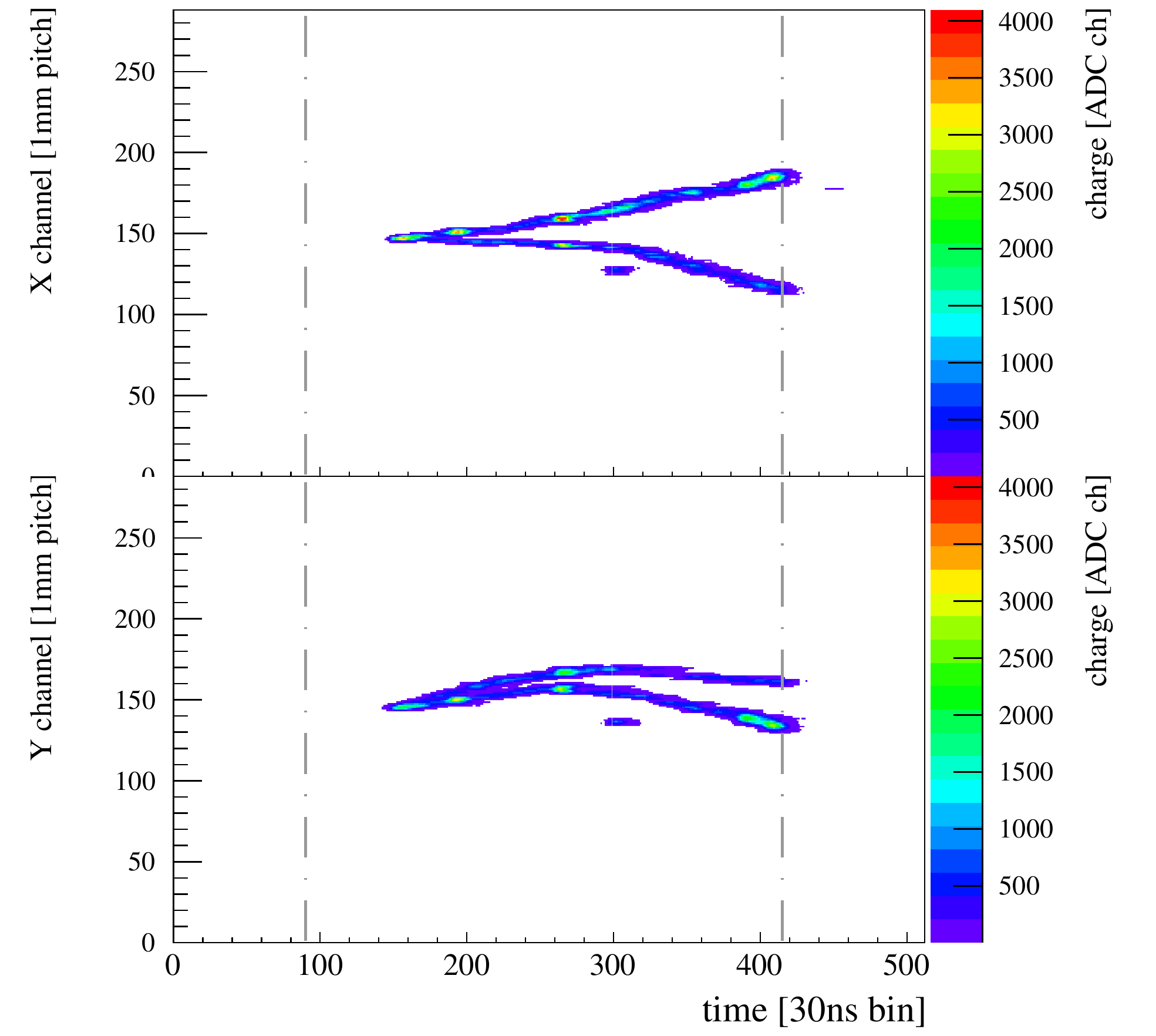}
      }
      \put(0.65,0.91){\normalsize Pair}
      \put(0.65,0.87){\normalsize 4.68\,\mega\electronvolt}
      \put(0.25,0){
        \includegraphics[width=0.5\unitlength]{\plotdir/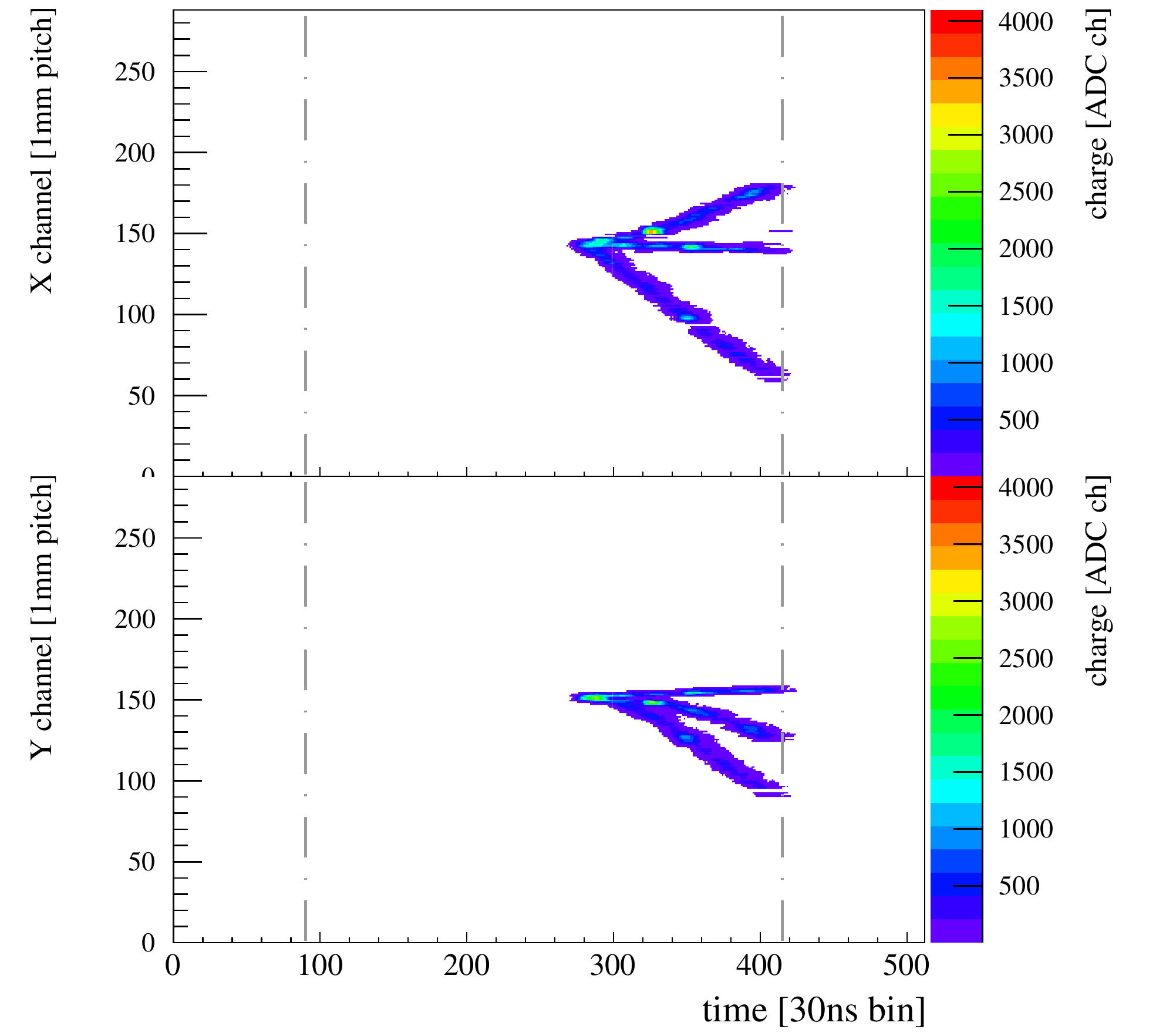}
      }
      \put(0.4,0.41){\normalsize Triplet}
      \put(0.4,0.37){\normalsize 74.3\,\mega\electronvolt}
    \end{picture}
    \caption{
      Example of events recorded in the HARPO detector.
      The three events are identified respectively as Compton scattering, pair conversion and triplet conversion.
      In the pair event, there is a small isolated energy deposit from interaction the following pulse (with 5\,\micro\second\ delay) on the readout plane.
      \label{fig:analyses:selection:events}
    }
  \end{center}
\end{figure}

Vertices in the beam region have a high probability to have originated from an interaction of a gamma ray with the gas.
These interactions are Compton scattering, pair production in the field of a nucleus (``pair'') or in the field of an electron (``triplet'').
They are distinguished by the opening angle $\theta_{+-}$.
Figure~\ref{fig:analyses:selection:events} shows examples of recorded events of each type.

\begin{figure}[ht]
  \begin{center}
    \iftrue
    \setlength{\unitlength}{0.94\textwidth}
    \begin{picture}(1,1)(0,0)
      \put(0,0.35){
        \includegraphics[width=0.5\unitlength]{\plotdir/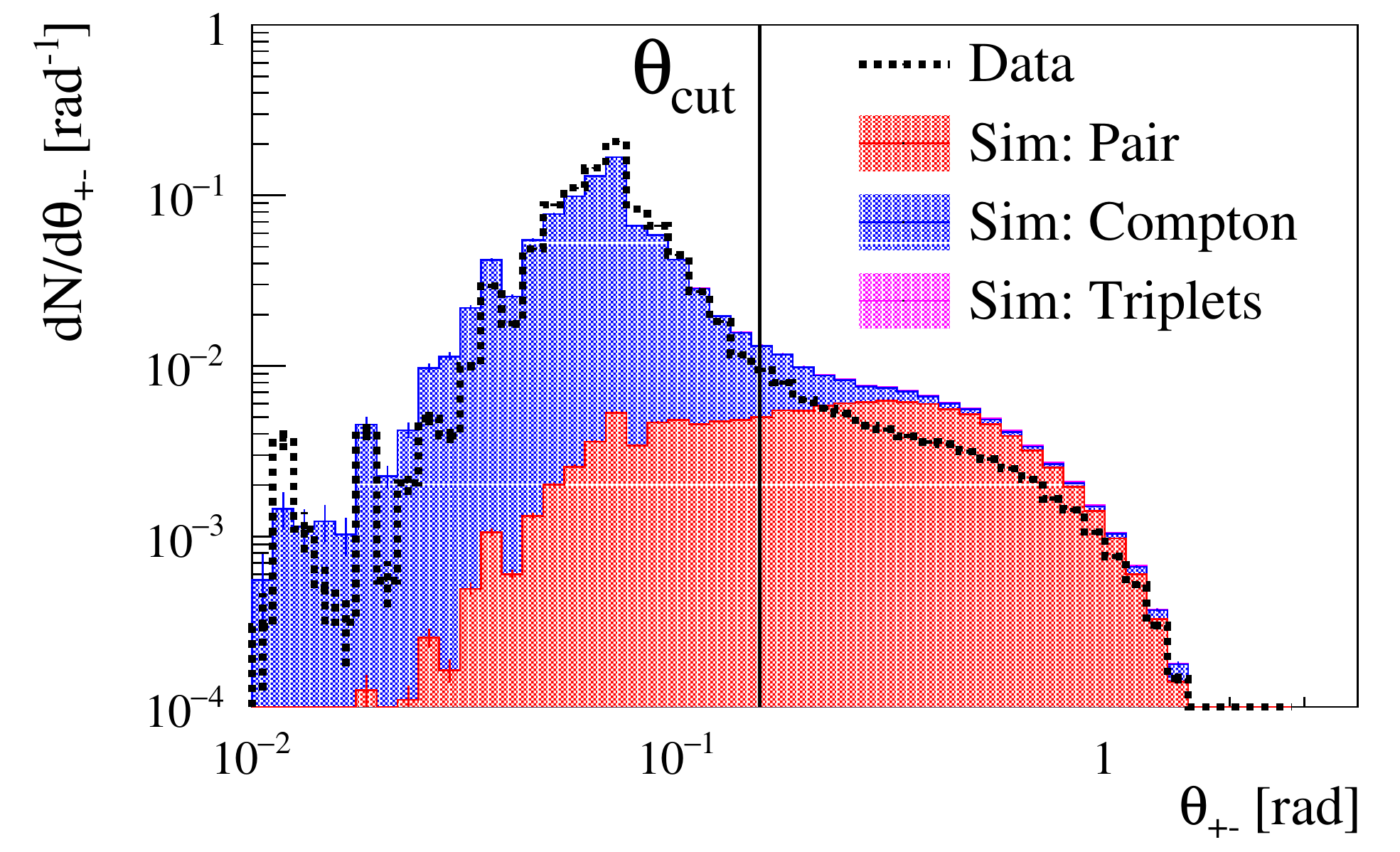}
      }
      \put(0.5,0.35){
        \includegraphics[width=0.5\unitlength]{\plotdir/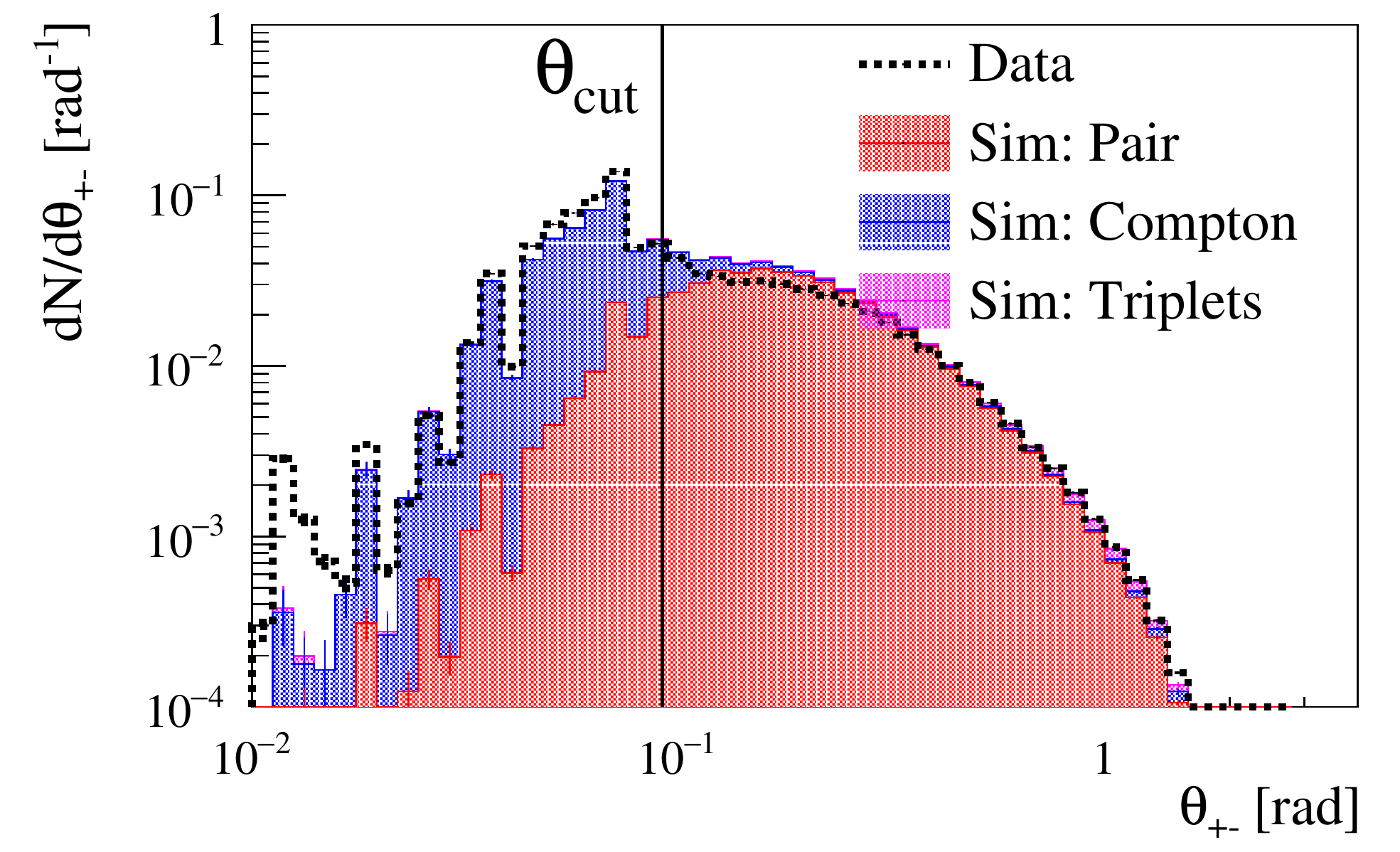}
      }
      \put(0,0){
        \includegraphics[width=0.5\unitlength]{\plotdir/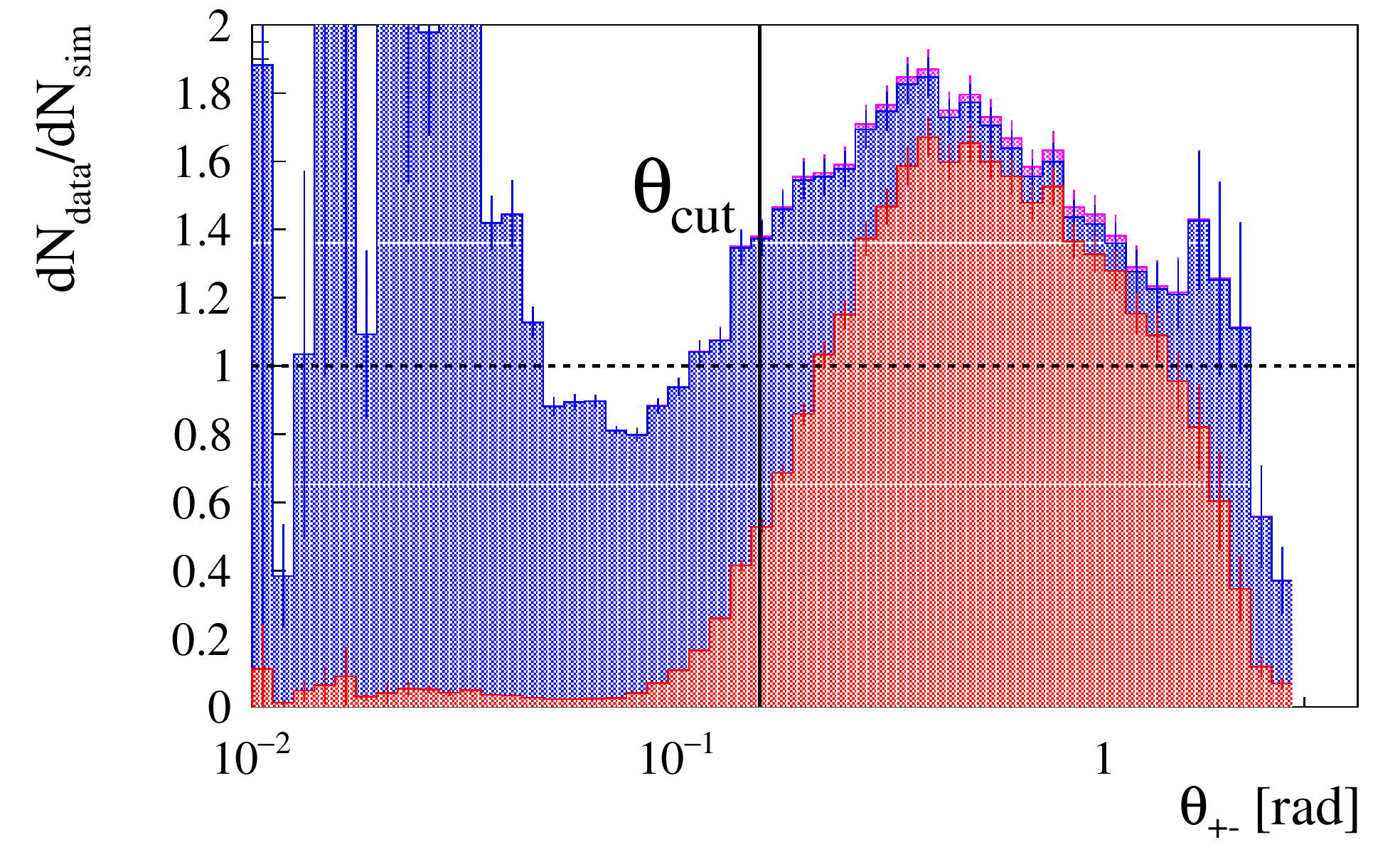}
      }
      \put(0.5,0){
        \includegraphics[width=0.5\unitlength]{\plotdir/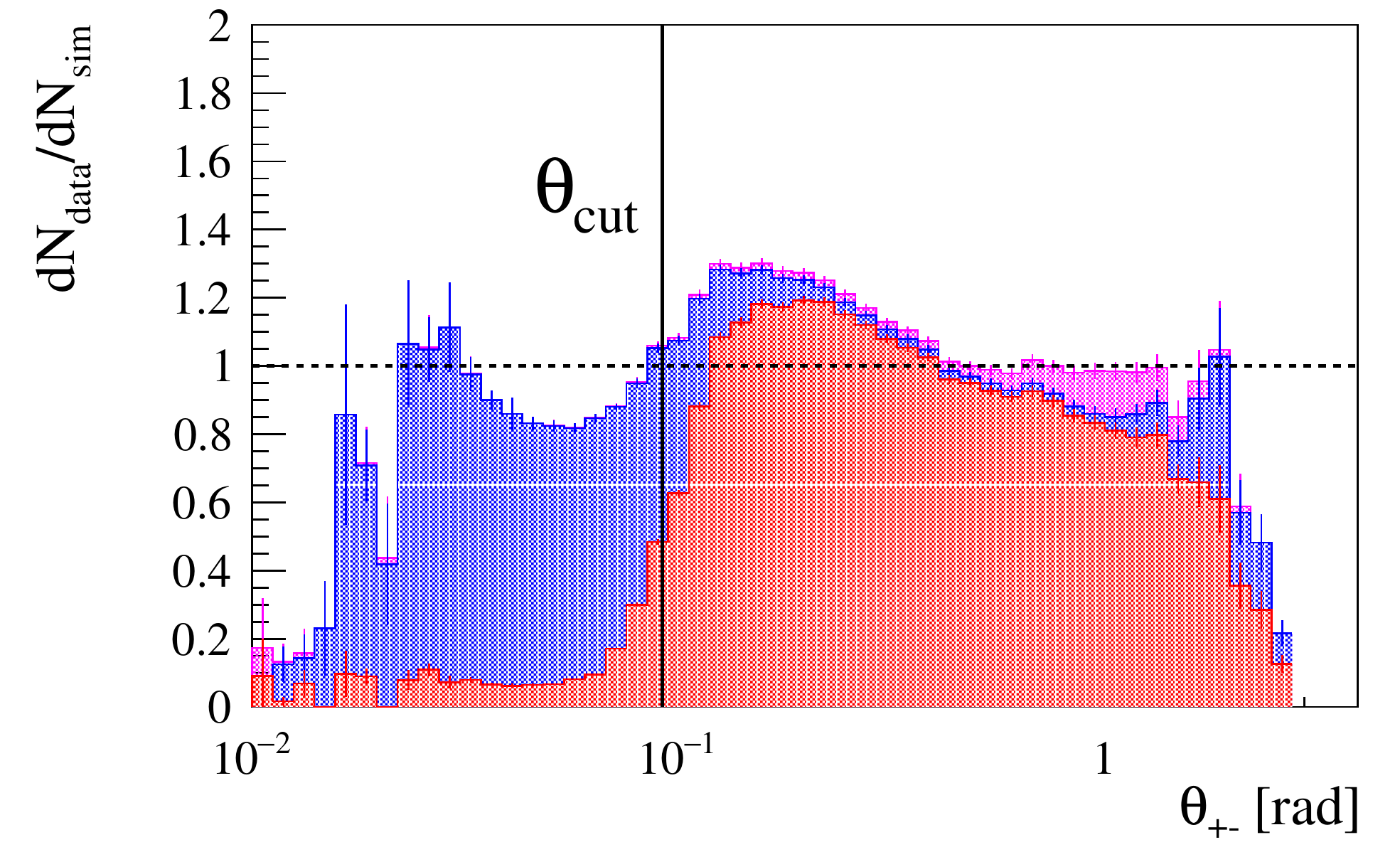}
      }
      \put(0.25,0.42){\large 4.68\,\mega\electronvolt}
      \put(0.75,0.42){\large 11.8\,\mega\electronvolt}
    \end{picture}
    \else
    \includegraphics[width=0.47\linewidth]{\plotdir/compareThetaPm_4MeV.pdf}
    \includegraphics[width=0.47\linewidth]{\plotdir/compareThetaPm_11MeV.pdf}
    \includegraphics[width=0.47\linewidth]{\plotdir/compareThetaPmRatio_4MeV.pdf}
    \includegraphics[width=0.47\linewidth]{\plotdir/compareThetaPmRatio_11MeV.pdf}
    \fi
    \caption{
      Distribution of the opening angle for 4.68\,\mega\electronvolt\ (left) and 11.8\,\mega\electronvolt\ (right) photon beam energies.
      The ratio between the data and the simulation is shown in the bottom row.
      At low energy, the proportion of pair production events seems overestimated in the simulation.
      \label{fig:analyses:selection:thetapm}
    }
  \end{center}
\end{figure}

For Compton scattering (single track), events can be misreconstructed as pair production (two track) events with small $\theta_{+-}$.
Figure~\ref{fig:analyses:selection:thetapm} shows the distribution of $\theta_{+-}$ for 4.68\,\mega\electronvolt\ and 11.8\,\mega\electronvolt\ beam energies, and compares them to simulations, decomposed into Compton scattering, pair and triplet events.
The angular distribution is well described by the simulation.
The proportion of Compton and pair events is however not correctly reproduced by the simulation at low energy, suggesting that the trigger efficiency is different for these two categories of events.

\begin{figure}[ht]
  \begin{center}
    \includegraphics[width=0.82\linewidth]{\plotdir/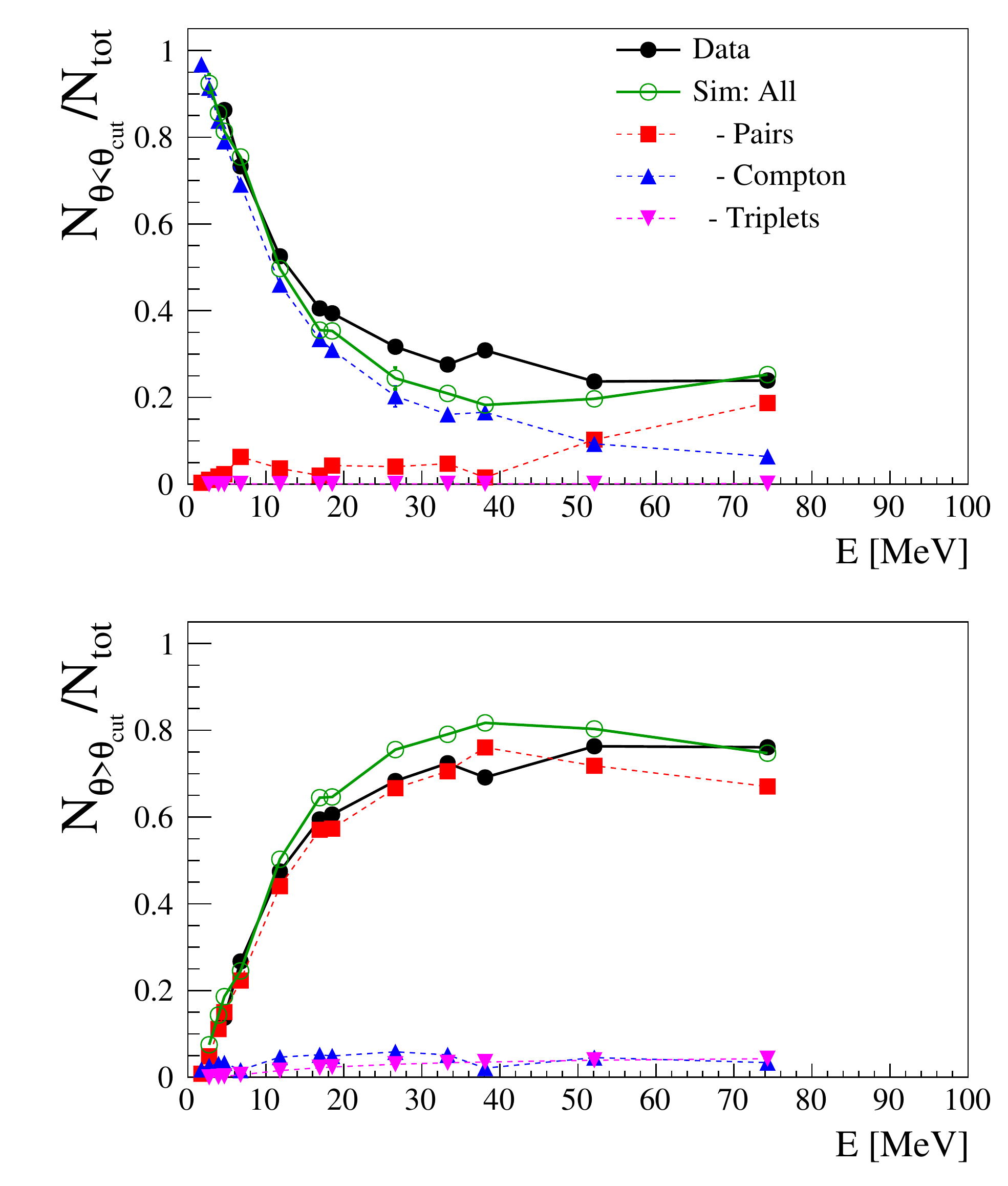}
    \caption{
      Fraction of the reconstructed vertices that are rejected (top) or selected (bottom) by the cut on the opening angle $\theta_{+-}$.
      The full lines show the total for data (in black) and simulation (in green).
      The dashed lines show the contributions of the different processes (Compton, pair and triplets).
      The data are well reproduced by the simulation, and above 4\,\mega\electronvolt\ more than 90\,\% of the selected vertices are from pair conversions.
      \label{fig:analyses:open:compareCutThetaPm}
    }
  \end{center}
\end{figure}

An energy-dependent cut on the opening angle $\theta_{+-}$ is used to select the pair events:
\begin{equation}
\theta_{+-} > \left(0.05+\frac{0.5\,\mega\electronvolt}{E_{\gamma}}\right)\,\radian.
\end{equation}
Figure~\ref{fig:analyses:open:compareCutThetaPm} shows the effect of the cut on the opening angle $\theta_{+-}$ on the different simulation samples (Compton, pair and triplet), and for the real data.
There is a good agreement between data and simulation.
The contamination of the pair production events by Compton scattering events is only of a few percent.
Above 40\,\mega\electronvolt, the opening angle gets smaller, and it is difficult to distinguish pair events from Compton scattering events.

The selection gives us high purity (over 90\%) pair event samples above 4\,\mega\electronvolt.
At low energy, the stronger cuts applied to remove Compton events reduce the reconstruction efficiency.
At high energy, the efficiency is affected by the low opening angles of the conversion events, which cannot be recognised as pairs.
In spite of these difficulties, simulations show that the vertices are reconstructed with an efficiency higher than 90\% over the whole spectrum presented here.

\section{Simulation of the HARPO detector}

The cubic geometry of the detector and the configuration of the readout scheme introduce a systematic bias to the polarisation measurement which cannot be addressed analytically.
It is therefore necessary to have an accurate simulation of the TPC.
We developed a complete simulation to describe the response of the HARPO detector.
It contains three main components: 
\begin{itemize}
\item An event generator describes the conversion of photons in the gas~\cite{Bernard:2013jea,Gros:2016zst}. It provides the energy momentum of the electron-positron pair.
\item The interaction of the electron and positron pair with the gas is simulated using Geant4~\cite{Geant4}. It provides the ionisation electrons in the gas volume.
\item The processes and the geometry of the TPC are described with a custom software~\cite{Gros:SPIE:2016} that provides a signal map similar to that of the real data.
\end{itemize}

The first two components have been validated in~\cite{Amako:2005xf} and~\cite{Gros:2016zst} respectively.
The last one was developed specifically for HARPO. 
The description of the TPC includes electron drift, diffusion and amplification in the gas, the readout space and time response, and the signal digitisation, including known electronics saturation effects.

\begin{figure} [ht]
  \begin{center}
    \includegraphics[width=0.75\textwidth]{\plotdir/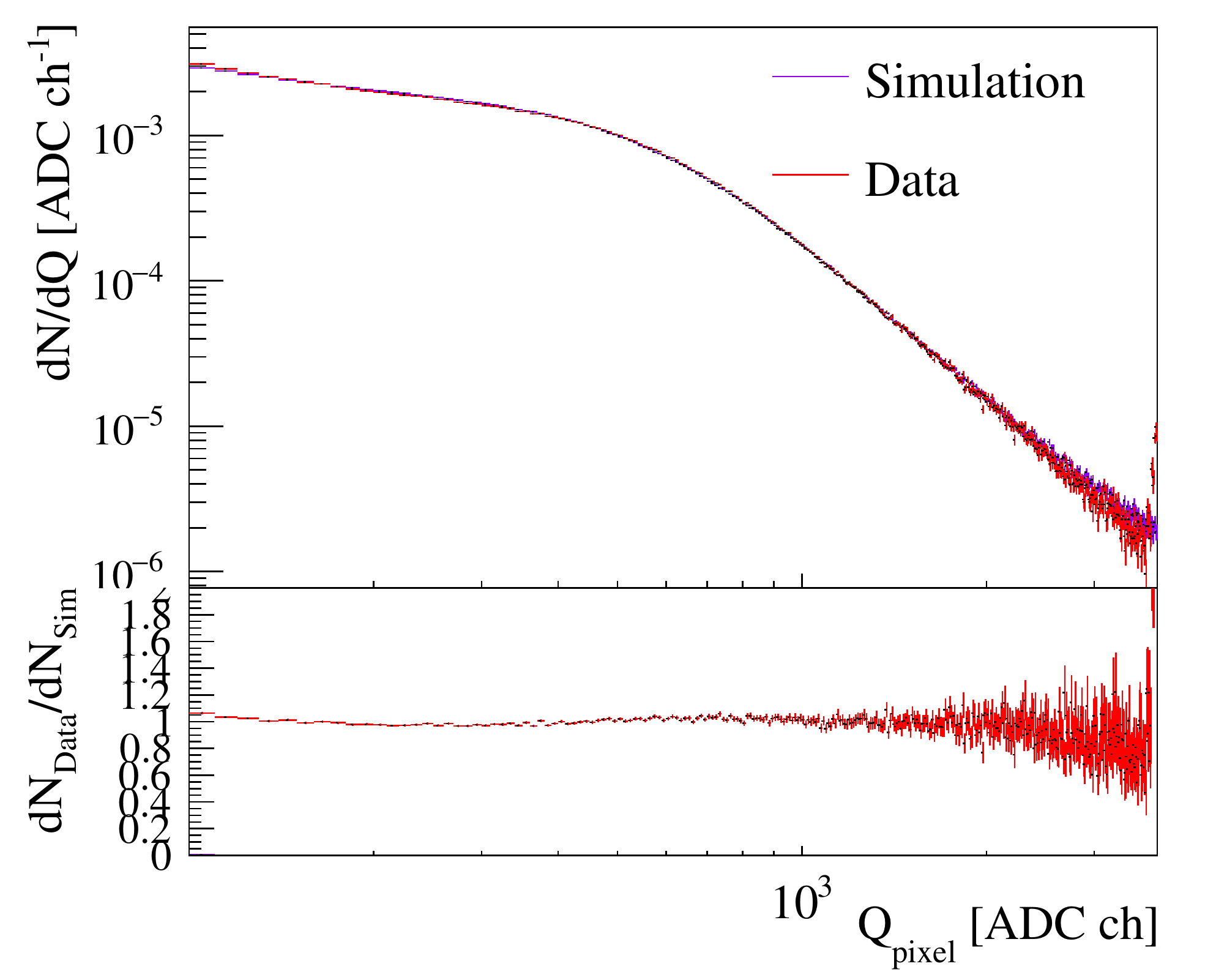}
  \end{center}
  \caption{
    \label{fig:calibQraw} 
    Comparison between data and simulation of the signal amplitude for single pixels.
    There is excellent agreement between the data and simulation.
  }
\end{figure}

This simulation was thoroughly validated using a tight selection of cosmic rays.
All of the simulation parameters were calibrated against data~\cite{Gros:TPC:2016}.
Figure~\ref{fig:calibQraw} shows an example of the comparison between cosmic-ray data and the simulation of the raw-charge read out for each channel and time bin.
This distribution is affected by most of the effects mentioned above.
There is an excellent agreement between data and simulation.

\section{Angular resolution\label{sect:analysis:gammaResolution}}

\begin{figure}[ht]
\begin{center}
  \includegraphics[width=0.82\linewidth]{\plotdir/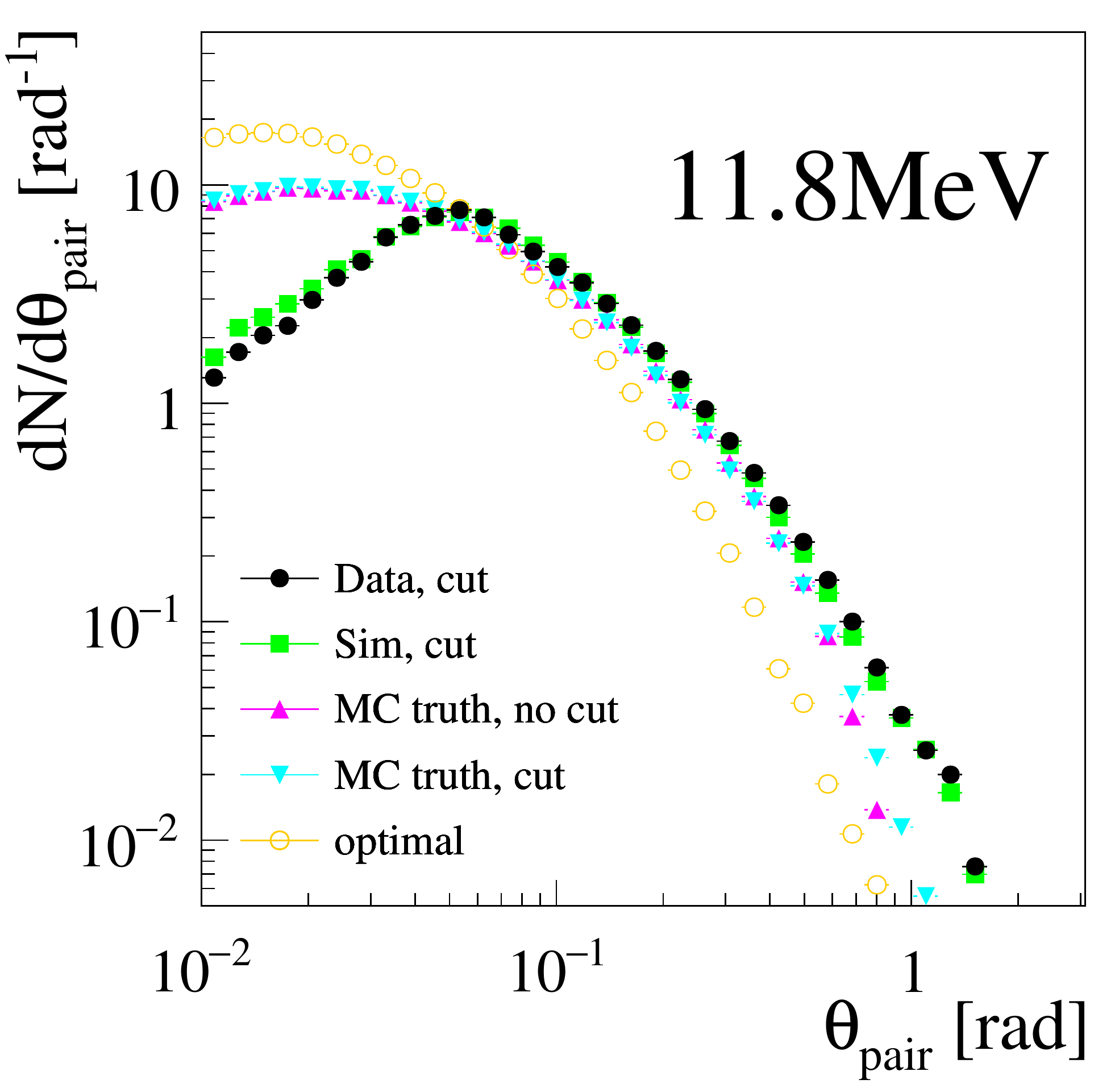}
  \caption{
    Distribution of the residual angle $\theta_{\rm pair}$ for a 11.8\,\mega\electronvolt\ photon beam.
    Black dots are the beam measurements.
    Yellow circles show the QED limit, using the full 4-vector information from the Monte-Carlo.
    The magenta triangles show the residuals using only the track direction from Monte-Carlo (unknown magnitude of the momentum), and the cyan triangles show the effect of the opening angle cut on that distribution.
    Finally, the green squares show the distribution taking into account the full detector simulation, including contamination from Compton and triplet events.
    The data are consistent with the full detector simulation.
    \label{fig:analyses:gammaResolution:thetapair}
  }
\end{center}
\end{figure}

From a reconstructed vertex, the corresponding photon direction is estimated as $\vec{u}_{\rm pair}=\vec{u}_{+}+\vec{u}_{-}$.
The residual angle $\theta_{\rm pair}$ is then:
\begin{equation}
  \theta_{\rm pair}=\arccos{(\vec{u}_{\rm pair}\cdot\vec{u}_{\gamma})},
\end{equation}
where $\vec{u}_{\gamma}$ is the beam direction.
After applying the vertex selection described in Sect.~\ref{sect:gammaSelection}, the distribution of the residual angle $\theta_{\rm pair}$ is obtained for each configuration of the same energy, polarisation and TPC orientation.
Such a distribution for 11.8\,\mega\electronvolt\ photons is shown in Fig.~\ref{fig:analyses:gammaResolution:thetapair}.

The angular resolution is dominated by three main effects:
\begin{itemize}
\item The momentum of the recoil nucleus is not measured.
The corresponding contribution is denoted $\sigma_{\rm recoil}$.
\item The magnitude of the momentum of the two particles is not measured.
The corresponding contribution is denoted $\sigma_{p}$.
\item The detector has a finite angular resolution for single  charged particles.
The corresponding contribution is denoted $\sigma_{\rm det,\gamma}$.
\end{itemize}
Figure~\ref{fig:analyses:gammaResolution:thetapair} shows the simulated distributions after neglecting each of these effects.
The full simulation (``Sim'') gives the resolution $\sigma_{\rm recoil}\oplus\sigma_{p}\oplus\sigma_{\rm det,\gamma}$.
Using the true track directions, the detector resolution is neglected (``MC truth''), and the resolution is $\sigma_{\rm recoil}\oplus\sigma_{p}$.
Using the full 4-vector information, the physical limit is reached (``optimal''), and the resolution is $\sigma_{\rm recoil}$.

\begin{figure}[ht]
\begin{center}
  \includegraphics[width=0.82\linewidth]{\plotdir/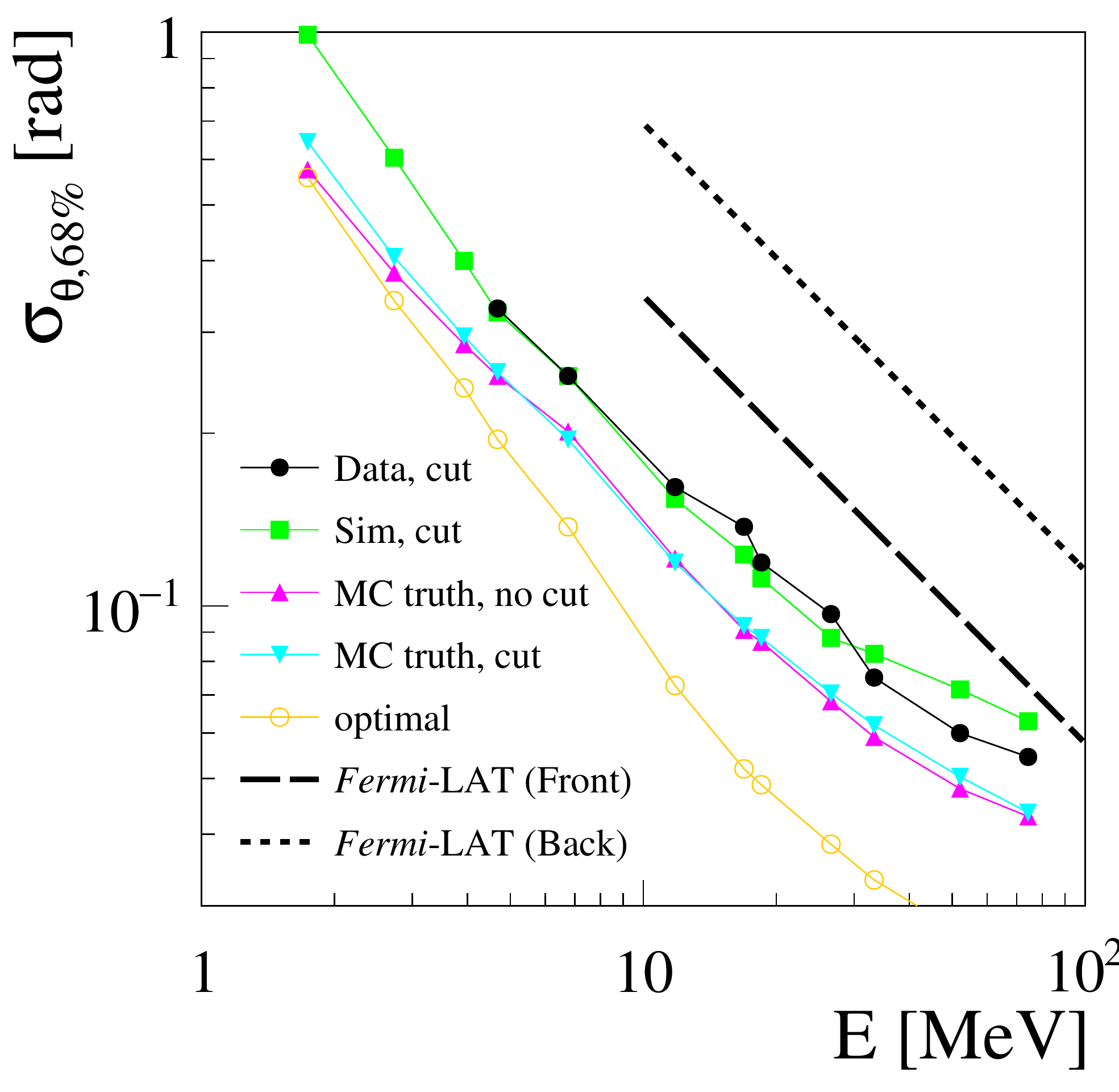}
  \caption{
    68\% containment angle as a function of the beam energy.
    The black dots show the measurements in beam.
    The green squares show the results of a full simulation, including contamination from Compton and triplet events.
    The magenta triangles show the QED limit in absence of momentum magnitude measurement (ignoring detector effects).
    The cyan triangles show the effect of the cut (in particular, on the opening angle) on the resolution, ignoring other detector effects.
    The yellow circles show the QED limit in absence of measurement of the recoiling nucleus.
    The angular resolution of \LAT\ for both front- and back-conversion events is shown for reference as the dashed and dotted lines.
    \label{fig:analyses:gammaResolution:resVsE}
  }
\end{center}
\end{figure}

The angular resolution $\sigma_{\theta,68\%}$ is defined as the 68\,\%~containment angle (i.e. the angle such that 68\,\% of the events have a smaller residual angle).
Figure~\ref{fig:analyses:gammaResolution:resVsE} shows the variation of the resolution with energy.
The beam data and the detector simulation are consistent.
The measured resolution is better by at least a factor of two than what is obtained by \LAT, even with a tracking-less reconstruction.
A large contribution to the resolution comes from the lack of momentum information.

\begin{figure}[ht]
\begin{center}
  \includegraphics[width=0.82\linewidth]{\plotdir/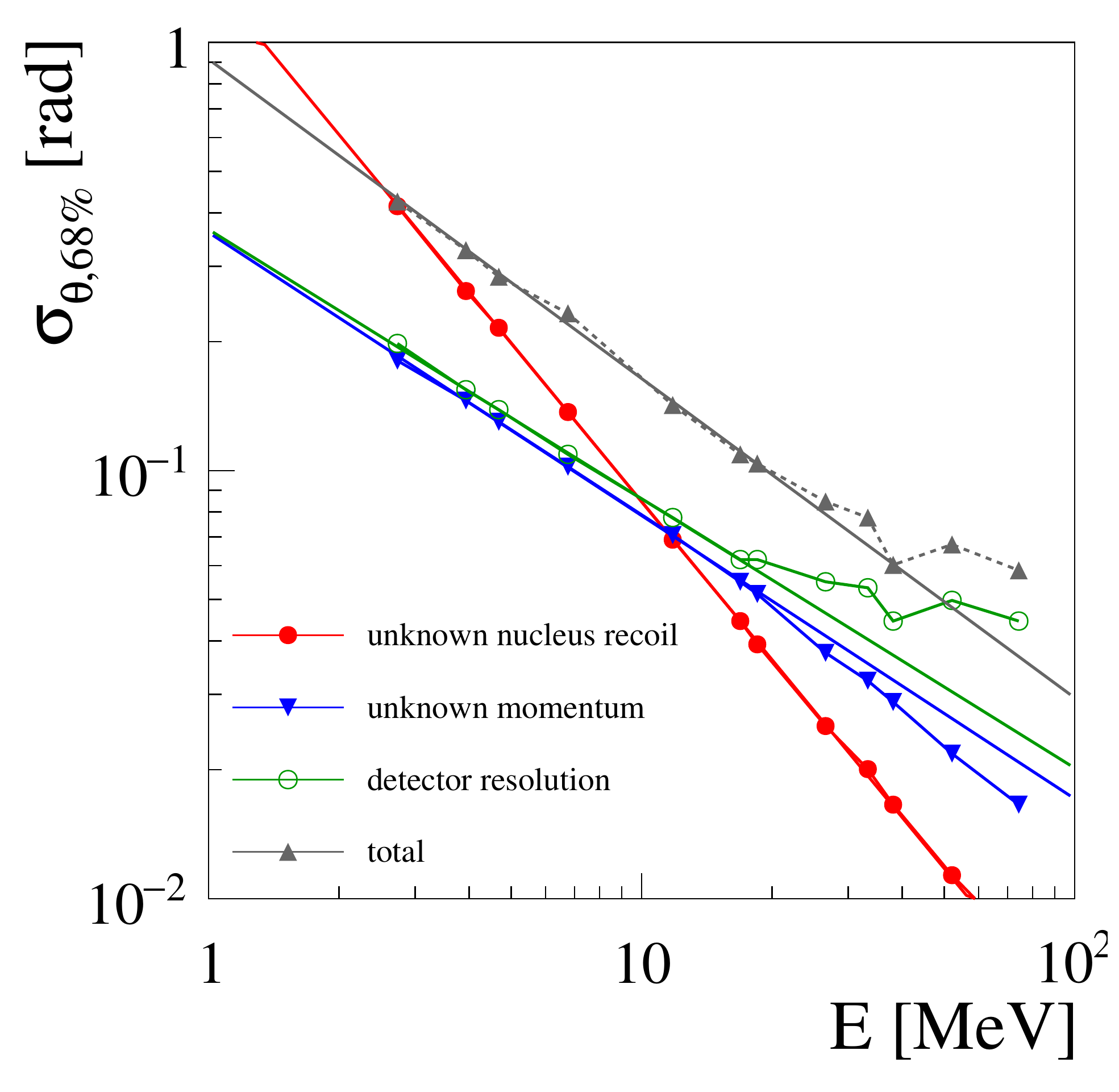}
  \caption{
    Break down of the contributions to the angular resolution for simulated photons converting with the pair process.
    Red filled circles show the effect of the unknown recoil momentum.
    Blue triangles show the effect of the unknown magnitude of the momentum of the two tracks.
    Green circles show the contribution of the detector effects, which are dominated by electronics saturation effects at high energy.
    This is fitted to a power law with index 0.64.
    The final resolution of the detector, in gray dashed line, is expected to be the quadratic sum of these components.
    Each graph is fitted with a power law shown as a straight solid line of the same color.
    This graph cannot be directly compared to the ``Sim, cut'' graph in Fig.~\ref{fig:analyses:gammaResolution:resVsE}, which also includes the contribution of Compton and triplet events.
    \label{fig:analyses:gammaResolution:resdiffVsE}
  }
\end{center}
\end{figure}

The three main components of the resolution ($\sigma_{\rm recoil}$, $\sigma_{p}$ and $\sigma_{\rm det,\gamma}$) are extracted from the simulation.
The results are shown in Fig.~\ref{fig:analyses:gammaResolution:resdiffVsE}.
Below 10\,\mega\electronvolt, the main contribution comes from the nucleus recoil.
The two other effects give a similar contribution.
The beam and detector geometry introduces a systematic bias at low energy (below 5\,\mega\electronvolt), so that the final resolution is not the quadratic sum of the components.
The two tracks in a pair are correlated, so that the actual resolution for single tracks is not relevant in this context.
An effective angular resolution for single tracks can be defined as $\sigma_{\rm det,e^{\pm}}=\sigma_{\rm det,\gamma}/\sqrt{2}$.
It can be approximated (see Fig.~\ref{fig:analyses:gammaResolution:resdiffVsE}) as:
\begin{equation}
  \sigma_{\rm det,e^{\pm}}\approx\left(\frac{p_{\pm}}{125\,\kilo\electronvolt/c}\right)^{-0.64}.
\label{eq:restrack}
\end{equation}

\clearpage

\section{Polarimetry}

Following~\cite{Gros:2016:azimuthal}, an optimal estimate of the polarisation asymmetry is given by the distribution of $\phi_{+-} = (\phi_{+}+\phi_{-})/2$, where $\phi_{\pm}$ is the azimuthal angle of the particle ($+$, positron; $-$, electron) with regard to the beam axis:
\begin{equation}
  \phi_{\pm} = \arctan\frac{u_{\pm}^X}{u_{\pm}^Y},
\end{equation}
where $X,Y$ are the coordinates in a plane perpendicular to the beam direction, and $X$ is the direction of the beam polarisation.

\begin{figure} [ht]
  \begin{center}
    \iftrue
    \setlength{\unitlength}{0.75\textwidth}
    \begin{picture}(1,1)(0,0)
      \put(0,0){
        \includegraphics[width=0.75\textwidth]{\plotdir/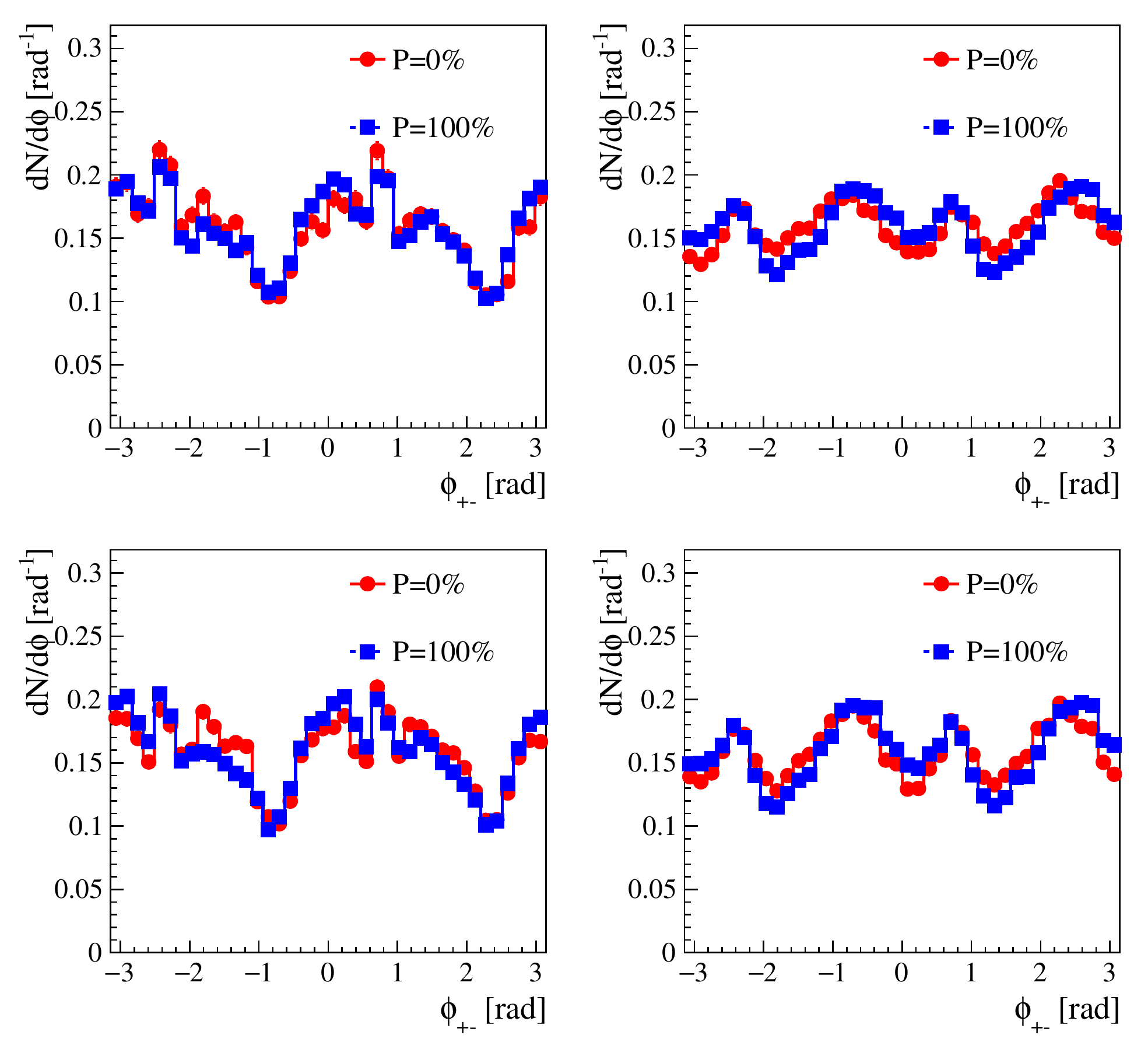}
      }
      \put(0.15,0.82){\large -45\degree}
      \put(0.65,0.82){\large 0\degree}
      \put(0.15,0.37){\large 45\degree}
      \put(0.65,0.37){\large 90\degree}
    \end{picture}
    \else
    \includegraphics[width=0.75\textwidth]{\plotdir/phipm_11MeV.pdf}
    \fi
  \end{center}
  \caption{
    \label{fig:analyses:polarimetry:phipm11MeV}
    Distribution of $\phi_{+-}$ for 11.8\,\mega\electronvolt\ photons, for four orientations of the HARPO TPC (-45\degree, 0\degree, 45\degree and 90\degree) around the beam axis.
    Data from polarised and unpolarised beam are shown with the blue squares and the red dots respectively.
  }
\end{figure}

The distribution of $\phi_{+-}$ is obtained for each configuration of same energy, polarisation and TPC orientation.
Figure~\ref{fig:analyses:polarimetry:phipm11MeV} shows an example of the distributions for 11.8\,\mega\electronvolt\ photons in each of the configurations.
There are large systematic effects due to the cubic geometry of the detector and the fixed direction of the photons.

\begin{figure} [ht]
  \begin{center}
    \iftrue
    \setlength{\unitlength}{0.75\textwidth}
    \begin{picture}(1,1)(0,0)
      \put(0,0){
        \includegraphics[width=0.75\textwidth]{\plotdir/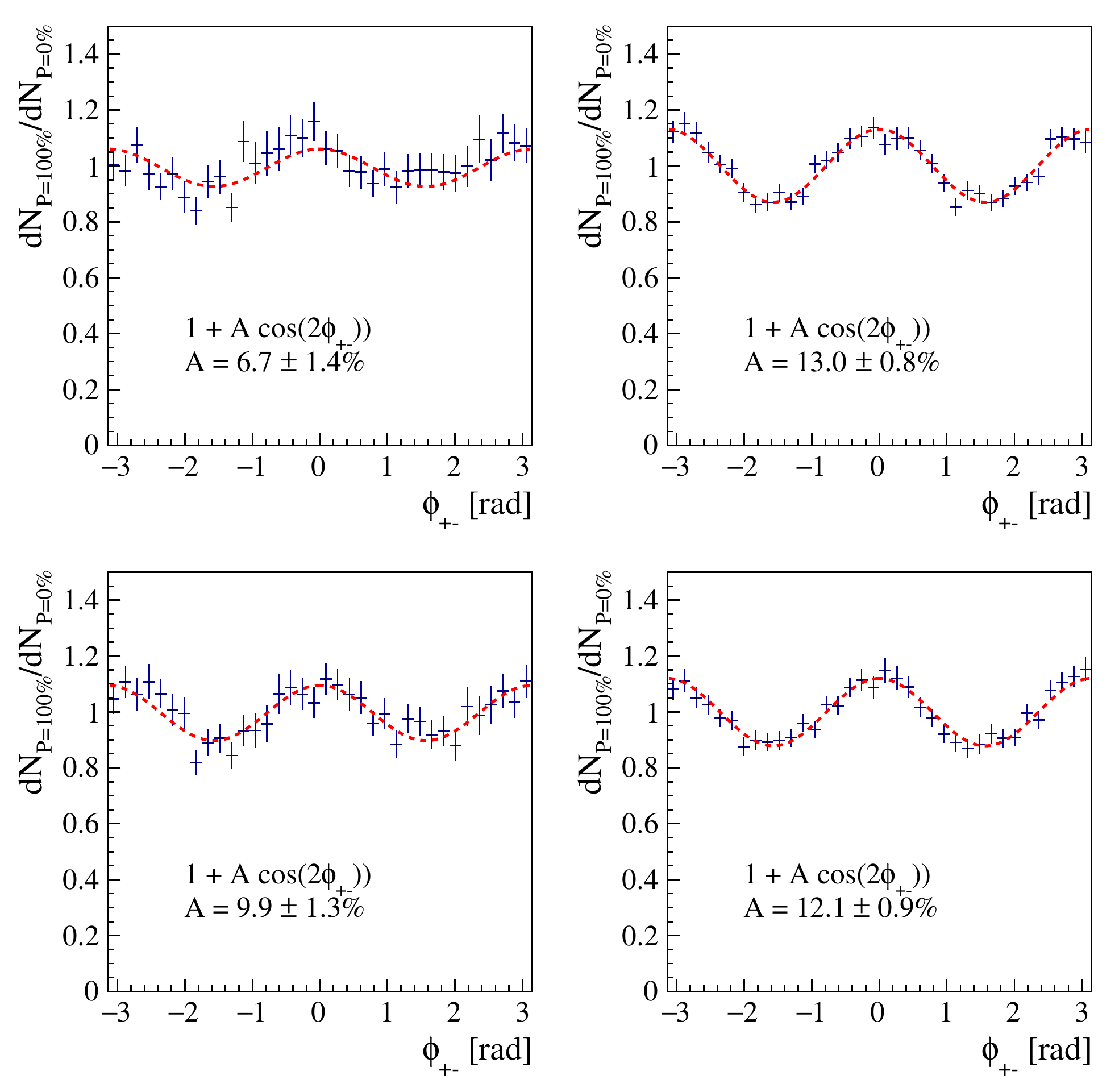}
      }
      \put(0.15,0.9){\large -45\degree}
      \put(0.65,0.9){\large 0\degree}
      \put(0.15,0.4){\large 45\degree}
      \put(0.65,0.4){\large 90\degree}
    \end{picture}
    \else
    \includegraphics[width=0.75\textwidth]{\plotdir/Apola_11MeV_0.pdf}
    \fi
  \end{center}
  \caption{
    \label{fig:analyses:polarimetry:example11MeV_angle}
    Ratio of the azimuthal angle distributions for polarised (P=100\%) and unpolarised (P=0\%) 11.8\,\mega\electronvolt\ photons.
    The systematic bias is cancelled by dividing the azimuthal angle distribution for polarised photons by the distribution for unpolarised photons (experimental data). 
    Four orientations of the detector around the beam axis (-45\degree, 0\degree, 45\degree, and 90\degree) were used.
  }
\end{figure}

\begin{figure} [ht]
  \begin{center}
    \iftrue
    \setlength{\unitlength}{0.94\textwidth}
    \begin{picture}(1,1)(0,0)
      \put(0,0.5){
        \includegraphics[width=0.5\unitlength]{\plotdir/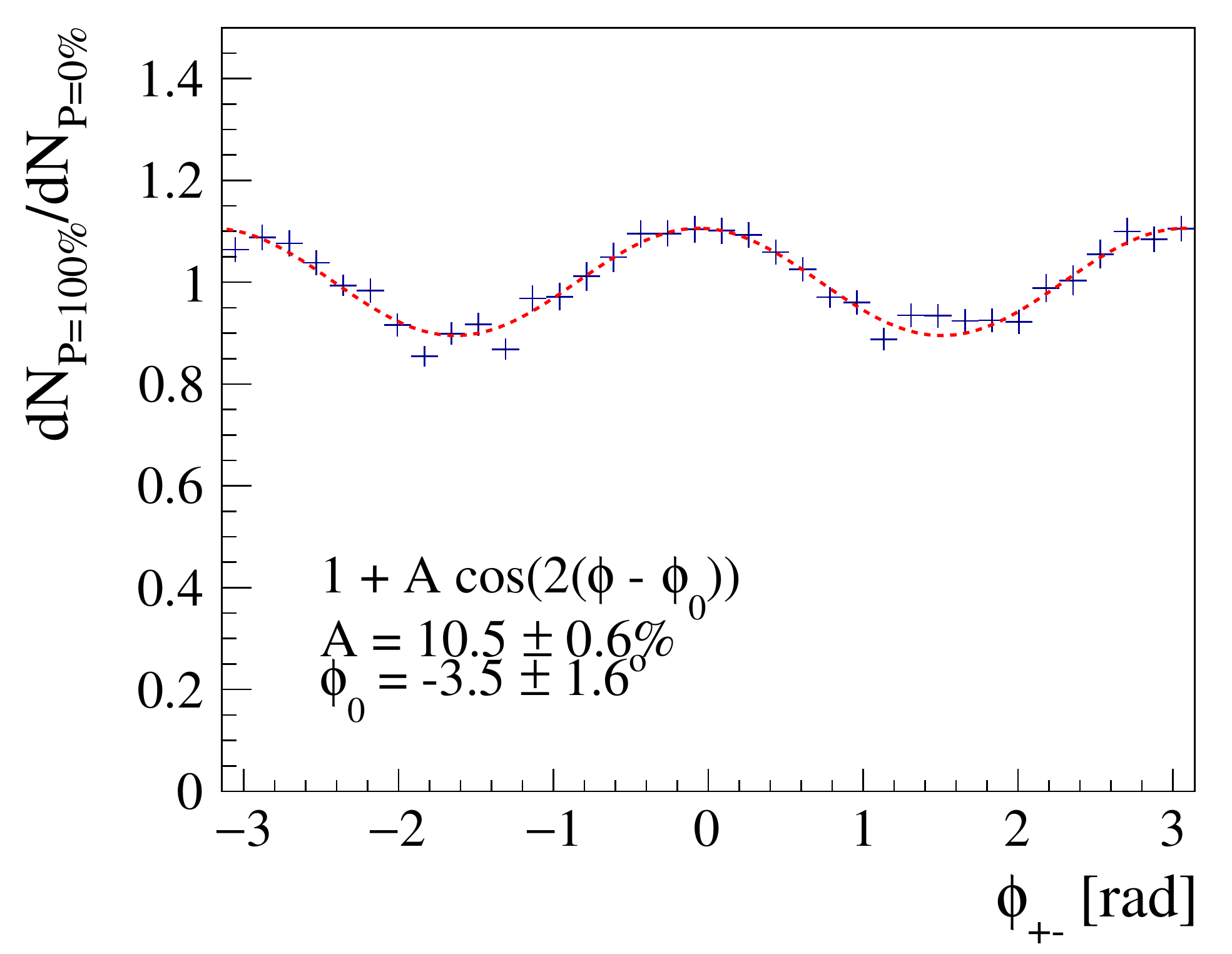}
      }
      \put(0.5,0.5){
        \includegraphics[width=0.5\unitlength]{\plotdir/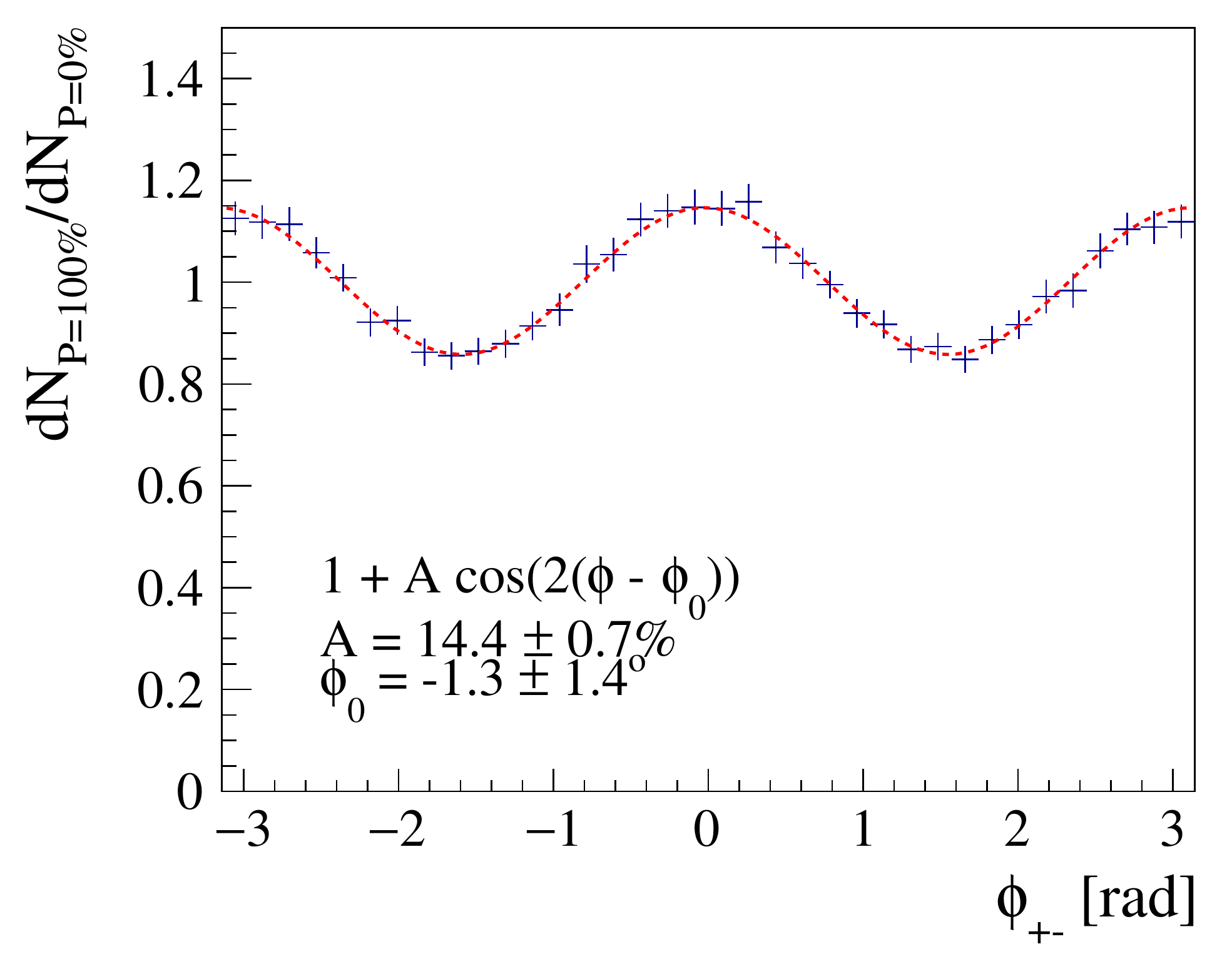}
      }
      \put(0.0,0.0){
        \includegraphics[width=0.5\unitlength]{\plotdir/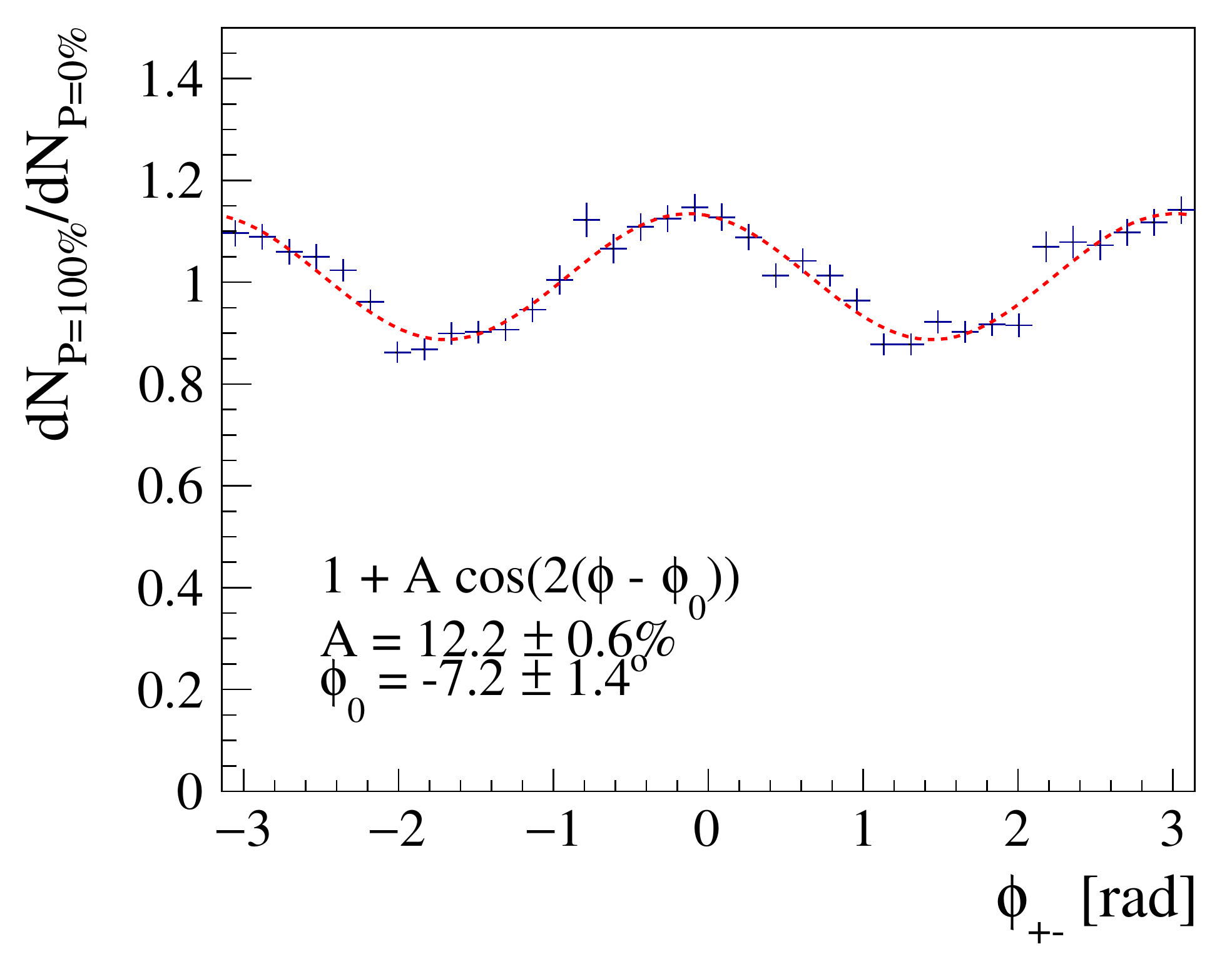}
      }
      \put(0.5,0.0){
        \includegraphics[width=0.5\unitlength]{\plotdir/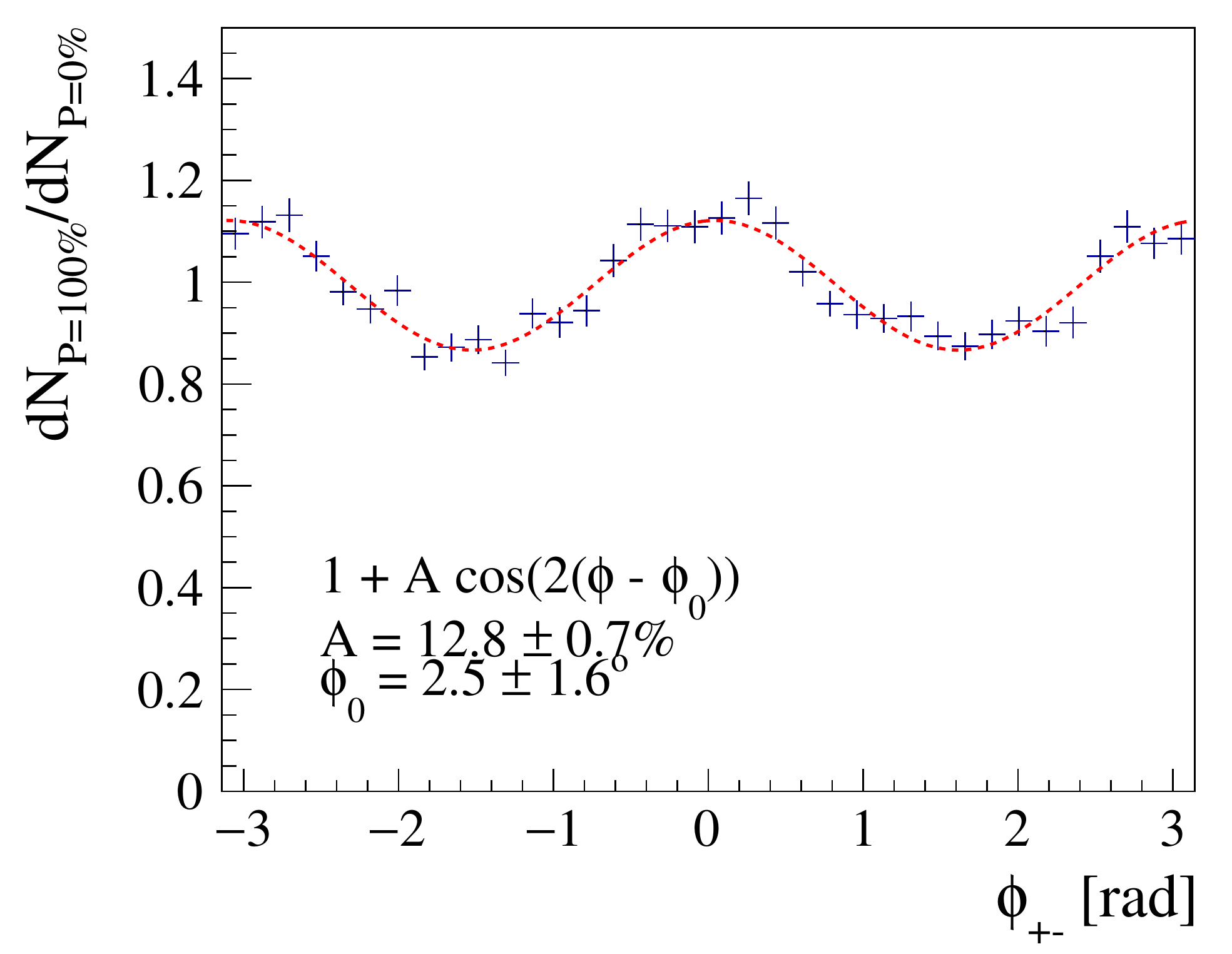}
      }
      \put(0.15,0.82){\large Data/Data}
      \put(0.65,0.82){\large Sim/Sim}
      \put(0.15,0.32){\large Data/Sim}
      \put(0.65,0.32){\large Sim/Data}
    \end{picture}
    \else
    \includegraphics[width=0.47\textwidth]{\plotdir/Apola_all_11MeV_0.pdf}
    \includegraphics[width=0.47\textwidth]{\plotdir/Apola_all_11MeV_11.pdf}
    \includegraphics[width=0.47\textwidth]{\plotdir/Apola_all_11MeV_1.pdf}
    \fi
  \end{center}
  \caption{
    \label{fig:analyses:polarimetry:example11MeV}
    Ratio of the azimuthal angle distributions for polarised (P=100\%) and unpolarised (P=0\%) 11.8\,\mega\electronvolt\ photons.
    Top left: ratio of polarised data over unpolarised data.
    Top right: ratio of polarised MC over unpolarised MC.
    Bottom left: ratio of polarised data over unpolarised MC.
    Bottom right: ratio of polarised MC over unpolarised data.
  }
\end{figure}

The geometry effects are cancelled out by taking the ratio of the polarised and unpolarised beam data.
Figure~\ref{fig:analyses:polarimetry:example11MeV_angle} shows the result for the four different orientations of the detector around the beam axis. 
The systematic bias is further reduced by combining the data with different orientations, resulting in Fig~\ref{fig:analyses:polarimetry:example11MeV}.
Since the unpolarised data are not available for every configuration, the simulation is used to correct the systematic bias.
Figure~\ref{fig:analyses:polarimetry:example11MeV}, bottom, shows the ratio of real polarised beam data with a simulated unpolarised beam.
In each case, the distribution is fitted with the expected function $1+A\cos{2(\phi_{+-}-\phi_{0})}$, where $A$ is the measured polarisation asymmetry.

\begin{figure} [ht]
  \begin{center}
    \includegraphics[width=0.75\textwidth]{\plotdir/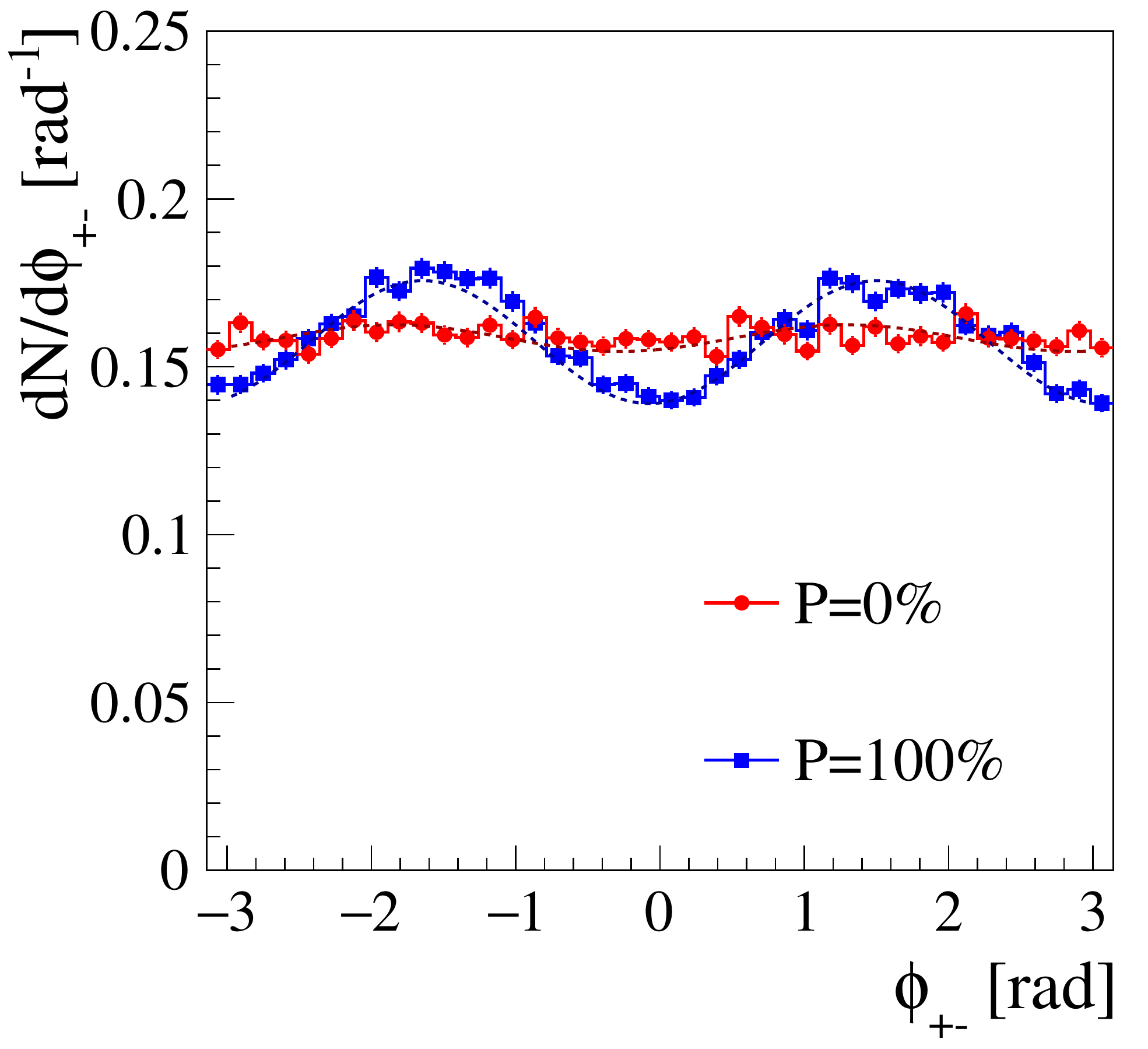}
  \end{center}
  \caption{
    \label{fig:analyses:polarimetry:phipm11MeVisotropic}
    Simulated distribution of $\phi_{+-}$ for 11.8\,\mega\electronvolt\ photons, for isotropic photons.
    The interaction points are uniformly distributed in the detector.
    Data from a polarised and an unpolarised source are shown with the blue squares and the red dots respectively.
  }
\end{figure}

The above results are strongly influenced by the fixed configuration of the photon beam.
In the case of a space telescope, the systematic bias would be very different.
Figure~\ref{fig:analyses:polarimetry:phipm11MeVisotropic} shows the azimuthal angle distribution for simulated isotropic 11.8\,\mega\electronvolt\ photons, converting uniformly inside the detector.
This represents a simple model for a long duration exposure in a space mission.
The reconstructed azimuthal angle is uniform for an unpolarised source, and shows the expected modulation when the photons are polarised.
In that case, no correction is applied.
The measured amplitude $A$ of the polarisation asymmetry in this case cannot be directly compared with what is measured in beam data.
The amplitude $A$ depends on the fiducial cuts chosen, because the angular resolution depends on the length of the particle trajectories inside the detector.
An optimisation of these cuts is necessary to assess the polarimetry potential in such a configuration.

\begin{figure} [ht]
  \begin{center}
    \includegraphics[width=0.75\textwidth]{\plotdir/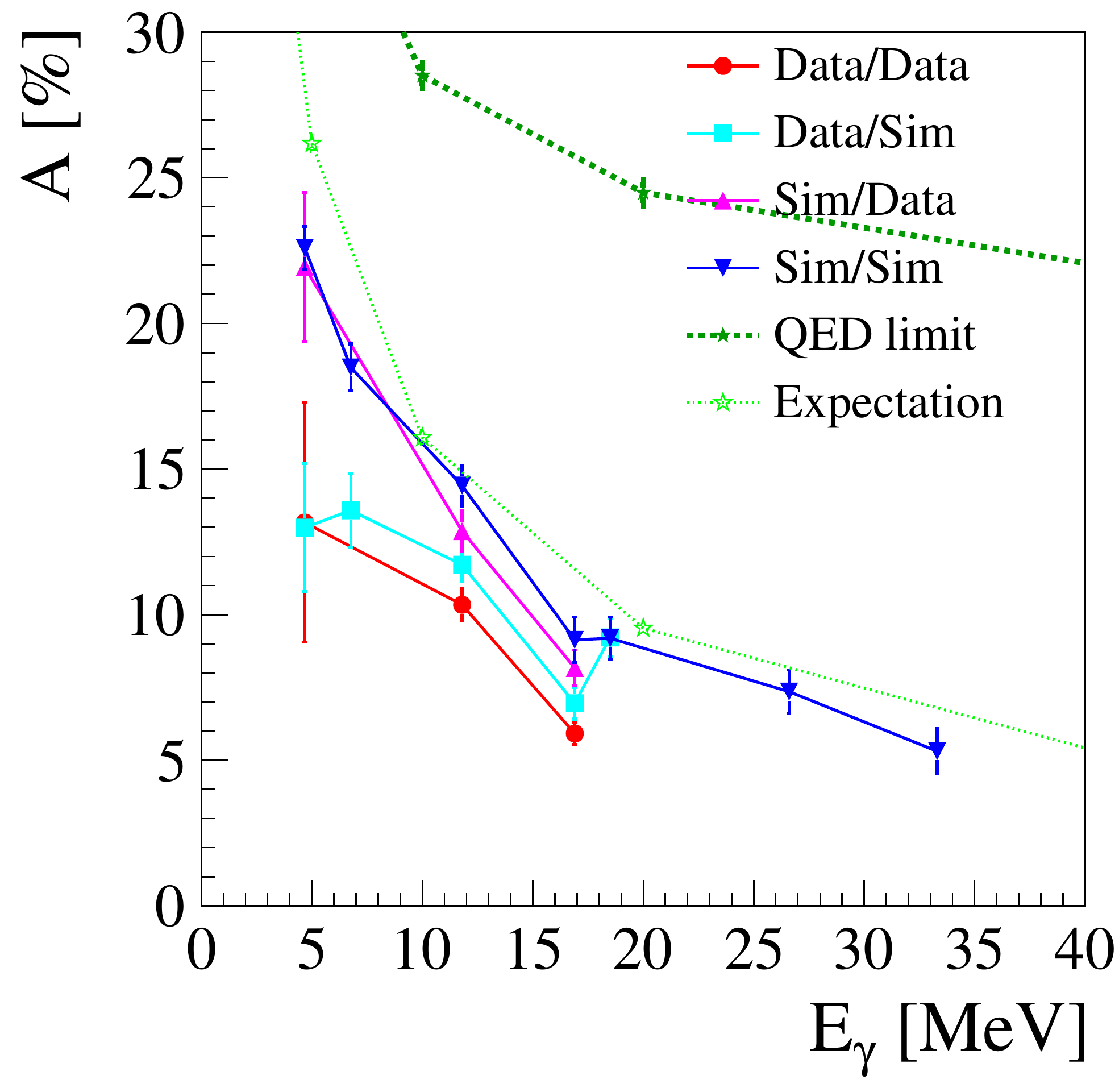}
  \end{center}
  \caption{
    \label{fig:analyses:polarimetry:ApolaVsE}
    Measured polarisation asymmetry as a function of the photon energy, using the ratio of $\phi_{+-}$ distributions for polarised and unpolarised photons.
    The QED limit from the pair-conversion kinematics is shown as a thick green dashed line, with full stars.
    The light green dashed line with open stars is this same limit, corrected to take into account the angular resolution for the tracks.
  }
\end{figure}

Figure~\ref{fig:analyses:polarimetry:ApolaVsE} shows the measured polarisation asymmetry $A$ obtained in each of the following cases:
\begin{itemize}
\item ratio of polarised data over unpolarised data;
\item ratio of polarised simulation over unpolarised simulation;
\item ratio of polarised data over unpolarised simulation;
\item ratio of polarised simulation over unpolarised data (as a validation of the method).
\end{itemize}
The optimal value of the polarisation asymmetry $A$ from QED is calculated using an exact event generator~\cite{Bernard:2013jea,Gros:2016:azimuthal}.
In addition to the full simulation represented in Fig.~\ref{fig:analyses:polarimetry:example11MeV}, we estimate the contribution of the single-track angular resolution, alone, to the dilution of the polarisation asymmetry.
The finite resolution on the azimuthal angle $\sigma_{\phi}$ dilutes this asymmetry by a factor $D=e^{-2\sigma^{2}_{\phi}}$.
Approximating the opening angle of the pair by its most probable value $\hat{\theta}_{+-} \approx E_{0}/E$ (with $E_{0}=$1.6\,\mega\electronvolt)~\cite{Olsen1963} gives:
\begin{equation}
\sigma_{\phi} \approx \frac{\sigma_{\theta,e^{+}} \oplus \sigma_{\theta,e^{+}}}{\hat{\theta}_{+-}} = \frac{\sqrt{2}\sigma_{\rm det,e^{\pm}}}{\hat{\theta}_{+-}}
\end{equation}
where $\sigma_{\rm det,e^{\pm}}$ is the effective angular resolution of the detector (Eq.~\ref{eq:restrack}).
This gives an expected value for the measured polarisation asymmetry, which is shown as the green dashed line and the open stars in Fig.~\ref{fig:analyses:polarimetry:ApolaVsE}.
The measured polarisation asymmetry is found to be consistent with the QED limit, taking into account the detector's angular resolution.

\clearpage

\section{Discussion}

The data shown here were obtained in a photon beam with high flux, which is not representative of the situation for a telescope for cosmic gamma rays.
This high event rate (up to 15\,\kilo\hertz, localised on the beam region of a few \milli\meter\squared) created extra difficulties related, in particular, to the saturation of the readout electronics.
In a space telescope, background tracks will be uniformly distributed in the gas volume, at a rate of about 20\,\kilo\hertz\per\meter\squared.
The event reconstruction could not be completed and optimised due to the limited resources of the project.
The results shown here should therefore be seen as a lower limit to the capabilities of the detector.

The detector performance can be further improved in several ways:
\begin{itemize}
\item reduction of the electronics saturation, by using a more adapted dynamic range;
\item improvement of the vertexing algorithm, using more adapted peak finding method;
\item introduction of a tracking algorithm, for event categorisation and possible resolution improvement;
\item introduction of a track-momentum estimation using multiple scattering\cite{Frosini:2017ft}.
\end{itemize}
All of these methods should improve the resolution of the detector, without requiring any hardware modifications.

The introduction of extra detectors to get acurate momentum measurement should of course be considered.
Conventional solutions such as a calorimeter or a magnet are however made very difficult by the weight constraints of a space telescope.

\section{Conclusion}

We have built a TPC for gamma-ray detection and polarimetry and we have successfully operated it in a polarised photon beam between 1.74\,\mega\electronvolt\ and 74\,\mega\electronvolt.
We have developed a simple but effective event reconstruction for pair production, and a detailed and accurate detector simulation.
Measurements of the angular resolution are consistent with simulations and offer an improvement of at least a factor of two over \LAT.
This resolution can be further improved with simple modifications of the hardware and software.
We show the first measurements of the polarisation of pair-production photons below 50\,\mega\electronvolt.

The behaviour of the detector is very well reproduced by the simulation.
We can now confidently use the simulation to investigate improvements, and to estimate the performance of such a detector for cosmic gamma-ray measurement in the high atmosphere or in space.

This work was funded by the French National Research Agency (ANR-13-BS05-0002) and was performed by using NewSUBARU-GACKO (Gamma Collaboration Hutch of Konan University).


\begin{thebibliography}{99}
 \small

\bibitem{McEnery:eASTROGAM}
 J. McEnery,
 ``Fermi-LAT below 100 MeV", 
``e-ASTROGAM workshop: the extreme Universe'', Padova Feb-March 2017

\bibitem{Bernard:2012uf}
 D.~Bernard [HARPO Collaboration],
 ``TPC in gamma-ray astronomy above pair-creation threshold,''
 Nucl.\ Instrum.\ Meth.\ A {\bf 701} (2013) 225.
 
\bibitem{Attie:2009zz}
 D.~Atti\'e,
 ``TPC review,''
 Nucl.\ Instrum.\ Meth.\ A {\bf 598} (2009) 89.
 
\bibitem{Bernard:2013jea}
 D.~Bernard [HARPO Collaboration],
 ``Polarimetry of cosmic gamma-ray sources above $e^+ e^-$ pair creation threshold,''
Nucl.\ Instrum.\ Meth.\ A {\bf 729} (2013) 765.

\bibitem{Moliere}
G. Moliere, 
``Theorie der Streuung schneller geladener Teilchen. III. Die Vielfachstreuung von Bahnspuren unter Ber\"uksichtigung der statistischen Kopplung'',
Zeitschrift Naturforschung A {\bf 10} (1955) 177.
 
\bibitem{Frosini:2017ft}
 M.~Frosini and D.~Bernard [HARPO Collaboration],
 ``Charged particle tracking without magnetic field: optimal measurement of track momentum by a Bayesian analysis of the multiple measurements of deflections due to multiple scattering,''
 Nucl.\ Instrum.\ Meth.\ A {\bf 867} (2017) 182.
 
\bibitem{Gros:TIPP:2014}
 P.~Gros {\em et~al.} [HARPO Collaboration],
 ``HARPO - TPC for High Energy Astrophysics and Polarimetry from the MeV to the GeV,''
 Proceedings of Science, {\bf TIPP2014} (2014) 133.

\bibitem{Frotin:2015mir}
 M.~Frotin {\it et al.} [HARPO Collaboration],
 ``Sealed operation, and circulation and purification of gas in the HARPO TPC,''
 arXiv:1512.03248 [physics.ins-det],
 to appear in EPJ Web of Conferences (MPGD2015)

\bibitem{GEM} 
 F.~Sauli,
``GEM: A new concept for electron amplification in gas detectors'',
 Nucl.\ Instrum.\ Meth.\ A {\bf 386}, 531 (1997).

\bibitem{micromegas}
 Y.~Giomataris {\em et~al.},
``MICROMEGAS: A high-granularity position-sensitive gaseous detector for high particle-flux environments'',
 Nucl.\ Instrum.\ Meth.\ A {\bf 376}, 29 (1996).

\bibitem{Bernard:2012jy}
 D.~Bernard [HARPO Collaboration],
``HARPO-A gaseous TPC for high angular resolution $\gamma$-ray astronomy and polarimetry from the MeV to the GeV,''
 Nucl.\ Instrum.\ Meth.\ A {\bf 718} (2013) 395.
 
\bibitem{Calvet2014zva}
 D.~Calvet,
 ``A Versatile Readout System for Small to Medium Scale Gaseous and Silicon Detectors,''
 IEEE Trans. Nucl. Sci. {\bf 61} (2014) 675.

\bibitem{Arogancia:2007pt}
 D.~C.~Arogancia {\it et al.},
 ``Study in a beam test of the resolution of a Micromegas TPC with standard readout pads,''
 Nucl.\ Instrum.\ Meth.\ A {\bf 602} (2009) 403.
 
\bibitem{Horikawa2010209}
 K.~Horikawa {\it et al.},
 ``Measurements for the energy and flux of laser Compton scattering $\gamma$-ray photons generated in an electron storage ring: NewSUBARU,''
 Nucl.\ Instrum.\ Meth.\ A {\bf 618 }(2010) 209.

\bibitem{Delbart:2015rmp}
 A.~Delbart [HARPO Collaboration],
 ``HARPO, TPC as a gamma telescope and polarimeter: First measurement in a polarised photon beam between 1.7 and 74 MeV,''
 PoS ICRC {\bf 2015} (2016) 1016.
 
\bibitem{Wang:TPC:2015}
 S.~Wang {\em et~al.} [HARPO Collaboration],
 ``HARPO: a TPC concept for $\gamma$-ray polarimetry with high angular resolution in the MeV-GeV range,''
J. Phys. Conf. Ser., {\bf 650} (2015) 012016.

\bibitem{Sun:2011es}
 C.~Sun and Y.~K.~Wu,
 ``Theoretical and simulation studies of characteristics of a Compton light source,''
 Phys.\ Rev.\ ST Accel.\ Beams {\bf 14} (2011) 044701
 
\bibitem{Geerebaert:2016dyv}
  Y.~Geerebaert {\it et al.} [HARPO Collaboration],
  ``Electronics for HARPO: Design, development and validation of electronics for a high performance polarised-Gamma-ray detector,''
 Real Time Conference (RT) (2016) IEEE-NPSS,
doi: 10.1109/RTC.2016.7543098
  
\bibitem{Wang:2015thesis}
 S.~Wang, ``Etude d'une TPC, cible active pour la polarim\'etrie et l'astronomie gamma par cr\'eation de paire dans HARPO'',
 PhD thesis, Ecole Polytechnique, 2015. {\em in French}

\bibitem{Gros:SPIE:2016}
P.~Gros {\it et al.} [HARPO Collaboration],
``Measurement of polarisation asymmetry for gamma rays between 1.7 to 74 MeV with the HARPO TPC'',
 {\bf SPIE2016} 9905-95.

\bibitem{Geant4}
 J.~Allison {\em et~al.},
 ``Geant4 developments and applications,''
 IEEE Transactions on Nuclear Science {\bf 53} (2006) 270.

\bibitem{Amako:2005xf}
 K.~Amako {\it et al.},
 ``Comparison of Geant4 electromagnetic physics models against the NIST reference data,''
 IEEE Trans.\ Nucl.\ Sci.\ {\bf 52} (2005) 910.

\bibitem{Gros:2016zst}
 P.~Gros and D.~Bernard [HARPO Collaboration],
 ``$\gamma$-ray telescopes using conversions to electron-positron pairs: event generators, angular resolution and polarimetry,''
 Astropart.\ Phys.\ {\bf 88} (2017) 60.

\bibitem{Gros:TPC:2016}
P.~Gros {\it et al.} [HARPO Collaboration],
``A TPC as high performance gamma-ray telescope and polarimeter: polarisation measurement in a beam between 1.7 and 74MeV with HARPO'',
proceedings to be published in Journal of Physics: Conference Series (JPCS).

\bibitem{Gros:2016:azimuthal} 
 P.~Gros and D.~Bernard [HARPO Collaboration],
 ``$\gamma$-ray polarimetry with conversions to e$^+$e$^-$ pairs: polarisation asymmetry and the way to measure it'',
 Astroparticle Physics {\bf 88} (2017) 30.

\bibitem{Olsen1963}
H. Olsen,
``Opening Angles of Electron-Positron Pairs'',
Phys. Rev. {\bf 131} (1963) 406.

\end{thebibliography}
\end{document}